\documentclass[12pt,a4paper]{amsart}
\usepackage{dsfont}

\title[Do Black Holes Exist in a Finite Universe Having the.... ]
{Do Black Holes Exist in a Finite Universe Having the Topology
of a Flat 3-Torus?}

\author{Frank Steiner }
\address{Institute of Theoretical Physics, Ulm University,  D--89069 Ulm, Germany}
\email{frank.steiner@uni-ulm.de}

\address{Univ Lyon, Ens de Lyon, Univ Lyon1, CNRS, Centre de Recherche Astrophysique de Lyon UMR5574, F--69007 Lyon, France}
\email{frank.steiner@ens-lyon.fr }

\keywords{General relativity: black holes, topology, large scale structure of universe, topological dark energy;  Appell and Epstein zeta function.}

\theoremstyle{definition}

\theoremstyle{remark}

\numberwithin{equation}{section}

\newcommand{\R}{{\mathbb R}}
\newcommand{\s}{{\mathbb S}}
\newcommand{\T}{{\mathbb T}}

\newcommand{\M}{{\mathbb M}}
\newcommand{\N}{{\mathbb N}}
\newcommand{\Z}{{\mathbb Z}}

\newcommand{\C}{{\mathbb C}}
\newcommand{\E}{{\mathbb E}}

\newcommand{\one}{{\mathds 1}}

\newcommand{\cJ}{{\mathcal J}}
\newcommand{\cL}{{\mathcal L}}

\DeclareMathOperator{\ttop}{top}
\DeclareMathOperator{\tot}{tot}

\DeclareMathOperator{\erfc}{erfc}
\DeclareMathOperator{\erf}{erf}

\def\vol{{\rm vol \,}}
\def\det{{\rm det \,}}

\def\Re{{\rm Re \,}}

\newcommand{\dt}{{\,dt}}
\newcommand{\ds}{{\,ds}}

\newcommand{\dx}{{\,dx}}
\newcommand{\dy}{{\,dy}}
\newcommand{\dz}{{\,dz}}

\begin{document}

\maketitle

\begin{abstract}\label{abstract}
Based on perturbation theory, we present the exact first-order solution to Einstein's gravitational field equations for the exterior static gravitational field of an isolated non-rotating star in a spatially finite universe having the topology of a flat 3-torus.  Since the method of images leads to a divergent Poincar\'e series, one needs a regularization which we achieve
by using Appell's resp. Epstein's zeta function. It turns out that the solution satisfies $R_{\mu\nu}=-\Lambda_{\ttop}g_{\mu\nu}$, where $\Lambda_{\ttop}$ is a new positive constant which is completely fixed by the mass $M$ of the star and the spatial volume $|\Omega|$ of the universe. The physical interpretation is that a stable or metastable equilibrium requires a {\em topological dark energy} which fills the whole universe with positive energy density $\varepsilon_{\ttop} \sim \Lambda_{\ttop}$ and negative pressure $p_{\ttop}=- \varepsilon_{\ttop}$.
The properties of the gravitational field are discussed in detail.
In particular, its {\em anisotropy} is made explicit by deriving an exact multipole expansion which shows that in this case Birkhoff's theorem does not hold. While the monopole describes the Newtonian potential, there is no dipole but always a non-vanishing quadrupole which leads to a repulsive force $\sim m \Lambda_{\ttop}|\vec{x}|$ experienced by a planet of mass $m \ll M$ at rest. Finally,  we put forward the {\em conjecture} that black holes exist in a toroidal universe and that their gravitational field is in the weak-field limit well approximated by the first-order field.
\end{abstract}

\section{Introduction} \label{intro}
Black holes are among the most fascinating astrophysical objects which were considered, however, for a long time as mere theoretical interpretations of some exact solutions to Einstein's field equations of General Relativity (GR)  [1 - 4].
Today there is no reasonable doubt \cite{5} that the compact radio source Sagittarius $A^\star, Sgr A^\star$, at the Galactic Centre is a supermassive black hole of about $4.4 \times 10^6M_{\odot}$.
Observations over a time span of now more than 2 decades of the proper motion of some stars (in particular the so-called $S$ stars) in the immediate vicinity of $ Sgr A^\star$ have shown that these stars move on Keplerian orbits around the central mass (see http://www.mpe.mpg.de/ir/GC for a 3D animation of the orbiting stars). Furthermore, observations and dynamical studies show that most galaxies contain near their centre a supermassive black hole of mass $10^6 - 10^9 M_{\odot}$. A prominent example is the giant elliptical galaxy M87 which contains at its centre a supermassive black hole of mass $(3.5 - 6.6)\times  10^9 M_{\odot}$ \cite{6}. Black holes like $Sgr A^\star$ and the one at the centre of M87 will play an important r\^{o}le in testing GR using for example the upcoming observations from the Event Horizon Telescope (EHT) (see http://www.eventhorizontelescope.org).\\

The recent detection of the transient  gravitational wave (GW) GW 150914 with the Advanced LIGO detectors \cite{7} is of the utmost importance for GR and for black hole physics in particular. First of all, this signal is the first direct detection of GWs predicted by Einstein exactly 100 years ago \cite{8, 9}. Secondly, the detected waveform matches the one predicted by GR for the inspiral and merger of a pair of black holes of mass $36^{+5}_{-4}M_{\odot}$ resp. $29^{+4}_{-4}M_{\odot}$
and the ringdown of the resulting single black hole of mass $62^{+4}_{-4}M_{\odot}$, with an energy $3.0^{+0.5}_{-0.5}M_{\odot}c^2$ radiated in GWs \cite{7}. This is also the first observation of a binary black hole merger and thus demonstrates the existence of binary stellar-mass black hole systems.
It is expected that many further GW signals will be observed during the coming years by the Advanced LIGO detector at Hanford, Washington and Livingstone, Lousiana in the US (www.ligo.org), the Advanced Virgo detector near Pisa in Italy (www.virgo-gw.eu) and the KARGA detector in the Kamioka-mine in Japan (gwcenter.icrr.u-tokyo.ac.jp/en). These observations will offer  the unique opportunity of testing GR with an unprecedented accuracy  in a regime where the theory has never been tested before. 
The detection of GW150914 has led to a vivid debate on the analysis and interpretation in the context of astrophysical predictions.
In addition various alternative scenarios are discussed, for example rotating gravastars as an alternative to black holes, naked singularities during the merger, firewalls, electromagnetic afterglows of GWs as a test of Quantum Gravity.\\

In the standard model of cosmology it is assumed that our universe possesses at large scales the space-time structure $\R \times \M^3$ where $\R$ describes cosmic time and $\M^3$ the 3-dimensional comoving spatial section of constant Gaussian curvature $K$. From measurements of e.g. the cosmic microwave background with NASA's satellite WMAP resp. ESA's satellite Planck one infers \cite{10,11} that $K$ is compatible with being zero and thus one concludes that $\M^3$ is flat, i.e. is a Riemannian manifold endowed with the euclidean metric. Since gravitation in GR is intimately linked to the geometry of space-time,  Einstein's original hope was  that his field equations would determine the global topology of space-time and thus also the spatial shape of the universe.
Having this in mind, he wrote his seminal paper of 1917 \cite{12} which laid the foundation of modern cosmology. In this paper he made ''the hypothesis of a spatially finite universe'' and identified $\M^3$ with $\s^3$ corresponding to constant positive curvature. In confronting this universe with the one of Newton where the 3-dimensional space is, of course, the infinite euclidean space $\E^3$, Einstein emphasized that in Newtonian theory one has to ''supplement the differential equations'' (Poisson's equation) ''by limiting conditions at spatial infinity'', but ''if it were possible to regard the universe as a continuum which is {\em finite (closed) with respect to its spatial dimensions}, we should not need at all of any such boundary conditons''. The Einstein universe is static which was compatible with all astronomical observations at the time. But in order to obtain a static universe, Einstein had to introduce the cosmological constant $\Lambda$ which he later abandoned when Hubble had discovered that our universe is expanding.
Today, a positive $\Lambda$ plays an important r\^{o}le in  the widely accepted $\Lambda C D M$ model, which assumes $\M^3=\E^3$ \cite{10,11}, in order to explain the accelerated expansion of the universe resp. the mysterious dark energy (see Sec. 3).\\

Since the gravitational field equations of GR are differential equations, they constrain the {\em local} properties of space-time but not the {\em global} spatial geometry and topology, i.e. they do not determine the spatial shape of the universe as originally hoped by Einstein. Thus ''his theory is incomplete'', since ''ignoring the global topology of the universe is particularly neglectful in a theory that purports to be a theory of geometry'' \cite{13}. It is likely that a future theory of Quantum Gravity will determine among other things also the curvature and topology resp. shape of the universe. In the meantime one attempts to infer the topology by confronting various models with astrophysical observations. This field
of research has been coined {\em cosmic topology} \cite{14} (see  also \cite{15} and references therein).\\

Taking the question of the global spatial topology of the universe serious, the first thing to know is which 3-dim. space forms are mathematically possible. Assuming that the spatial curvature at large is zero (which is compatible with present observations), we have to find all admissible 3-dim. euclidean space forms. Fortunately enough, this problem has already been solved a long time ago in connection with the 3-dim. crystallographic groups. It turns out \cite{13,14,15} that there are 18 possible space forms, where one is the infinite simply connected space $\E^3$ which is considered in the standard model of cosmology. Among the remaining 17 space forms there are 8 non-orientable manifolds.
Setting aside the non-orientable space forms, one is left with 9 orientable multiply connected manifolds of which 6 are compact, i.e. have a finite volume. It turns out  that the simplest of the orientable compact flat manifolds is the torus $\T^3$ also called hypertorus. Note that all of the 6 compact manifolds are globally anisotropic and all except of $\T^3$ destroy global homogeneity.\\

In this paper we  are not concerned with the time-evolution of the universe. Rather we are interested in the question whether there exist black hole solutions to Einstein's field equations in a universe whose spatial section is a 3-torus.
Since black holes are ideal objects to test GR, as we have seen before, they could give important information on the global topology of the universe, in particular on the question whether we live in a finite or infinite universe. Using a perturbative approach, we give an exact solution to the first-order Einstein equations in terms of Appell's \cite{16,17,18} and Epstein's \cite{19,20} zeta function. Based on this solution, we {\em conjecture} that there exist exact solutions describing black holes whose properties are very similar to those of the presented first-order solution.

\section{The Geometry of the Torus Universe} \label{sec2}

To describe the geometry of a flat 3-torus $\T^3$, it is convenient to use a right-handed rectangular coordinate system such that an arbitrary point in $\E^3$ is represented in terms of its cartesian coordinates by a 3-vector, e.g.
$\vec{a}=(a_1, a_2,a_3)$ with $a_k \in \R$. The scalar product of two vectors $\vec{a}$ and $\vec{b}$ is defined as usual,
$\vec{a} \cdot \vec{b}:=a_1 b_1 + a_2 b_2+a_3 b_3$, and the length of a vector $\vec{a}$ is $|\vec{a}|:=(\vec{a}{^2}){^{1/2}}\equiv (\vec{a}\cdot \vec{a}){^{1/2}}$. The euclidean distance between two points $\vec{x},\vec{y} \in \E^3$ is then given by $d(\vec{x},\vec{y}):=|\vec{x}-\vec{y}|$.\\

A general flat 3-torus $\T^3$ can be defined by a { \em fundamental domain} $\Omega \in \E^3$ having the shape of a parallelepiped (PEP) with 12 edges, 8 vertices and 6 faces (parallelograms). By identifying (''gluing'') opposite faces by pairs according to a discrete and fixed-point free symmetry group $\Gamma$, we obtain a multiply connected flat manifold having the topology of $\T^3$. Although the gluing cannot be performed in three dimensions, there is a neat representation on the covering space $\E^3$ based on the ''method of images''. By considering the set of all $\Gamma$-copies (congruent ''mirror images'') of $\Omega$, one obtains a tiling of $\E^3$,  whereby the set of all vertices of the PEPs defines the lattice points of a lattice $\cL$ in $\E^3$. The group $\Gamma$ is then given by one of the well-known three-dimensional crystallographic groups, and the corresponding torus $\T^3$ is the quotient $\E^3 /~ \Gamma$.\\

Any PEP can be completely characterized by a real positive-definite (3 x 3)-matrix $A=(a_{k\ell})_{k,\ell=1,2,3}$ which (without 
loss of generality) can be assumed to be given in lower triangular form with $a_{11}, a_{22}, a_{33} > 0, a_{21}, a_{31}, a_{32} \in \R$ and $a_{12}=a_{13}=a_{23}=0$. Let $\vec{a}_k=(a_{k1},a_{k2},a_{k3}), k=1,2,3,$ be three vectors  which are given by the rows of $A$. Then they define three edges of a PEP (which meet at the origin $\vec{0}$) with edge length $L_k:=|\vec{a}_k|,k=1,2,3$. Furthermore, the PEP corresponding to $A$ defines a {\em fundamental domain} $\Omega=\Omega(A)$ of a 3-torus $\T^3=\T^3 (A)$ with {\em volume}
\begin{equation} \label{equ2.1}
\vol (\T^3):=|\Omega|=(\vec{a}_1 \times \vec{a}_2)\cdot \vec{a}_3 = \det A=a_{11}a_{22}a_{33} \ .
\end{equation}

Obviously, the collection of all 3-tori can be represented as points in a 6-dimensional Teichm\"uller space. The symmetry group $\Gamma=\Gamma(A)$ associated with $A$ is defined by three generators (and their inverses) corresponding to the following discrete translations on the covering space $\E^3$
\begin{equation} \label{equ2.2}
\vec{x} \longmapsto \vec{x} \pm \vec{a}_k \quad (k=1,2,3) \ .
\end{equation}

Although the tori have no boundary, the transformations (\ref{equ2.2}) can be seen as {\em periodic boundary conditions} on $\partial \Omega$, and thus the basis vectors $\vec{a}_k$ play the r\^{o}le of {\em period vectors} resp. their coordinates define three {\em groups of periods} \cite{16}.\\

Let the vectors $\vec{b}_k=(b_{k1}, b_{k2}, b_{k3}) (k=1,2,3)$ be given by the columns of $A$, i.e. $b_{k\ell}:=a_{\ell k}$ and define the vector
\begin{equation} \label{equ2.3}
\vec{\rho}=\vec{\rho}(\vec{n}):=(\vec{n}\cdot \vec{b}_1,\vec{n}\cdot \vec{b}_2, \vec{n}\cdot \vec{b}_3) \ ,
\end{equation}
where $\vec{n}=(n_1,n_2,n_3)\in \Z^3$. Then $\vec{\rho}$ represents an arbitrary lattice point of $\cL$ located a distance $\rho=\rho (\vec{n}):=|\vec{\rho}(\vec{n})|$ from the origin. Note that $\rho^2$ is a quadratic form,
\begin{equation} \label{equ2.4}
\rho^2=\hat{Q}[\vec{n}]:=\sum\limits^3_{k,\ell=1}n_k\hat{Q}_{k\ell}n_\ell \ ,
\end{equation}
where $\hat{Q}:=AA^T$ is the Gram matrix corresponding to the period vectors $\vec{a}_k$.\\

From (\ref{equ2.3}) we  deduce $\vec{\rho}(\vec{0})=\vec{0},\vec{\rho}(-\vec{n})=-\vec{\rho}(\vec{n})$ and $\vec{\rho}(\vec{m})+\vec{\rho}(\vec{n})=\vec{\rho}(\vec{m}+\vec{n})$. Then the {\em group action} of any group element $\gamma_{\vec{n}} \in \Gamma$ acting on $\vec{x} \in \E^3$ is given by
\begin{equation} \label{equ2.5}
\vec{x}\longmapsto \gamma_{\vec{n}}(\vec{x}):=\vec{x}+\vec{\rho}(\vec{n}) \quad \forall \ \vec{n} \in \Z^3 \ .
\end{equation}
Since $\gamma_{\vec{m}} (\gamma_{\vec{n}}(\vec{x}))=\gamma_{\vec{m}+\vec{n}}(\vec{x}),\Gamma$ is an abelian group.
It is important to note that the manifold $\T^3$ is globally homogeneous, which implies, for example, that the measurements of the statistical properties of the cosmic microwave background do not depend on the position of the observer. However, $\T^3$ is not globally isotropic, since the global periodic identifications (\ref{equ2.2}) resp. (\ref{equ2.5}) break the rotational SO(3)-symmetry (even though they leave the flat metric on $\E^3$ invariant).\\

Scalar fields $\Phi(\vec{x})$ on $\T^3$, like the static gravitational field of a star or a black hole, typically inherit the symmetries of $\T^3$ and thus have to be {\em triply-periodic functions} satisfying the {\em periodicity conditions} [16, 17, 18]
\begin{equation} \label{equ2.6}
\Phi (\vec{x}+\vec{\rho}(\vec{n}))=\Phi(\vec{x})\quad \forall \ \vec{x} \in \E^3, \vec{n} \in \Z^3 \ .
\end{equation}
These functions are a natural generalization of  the well-known doubly-periodic functions, in particular of the meromorphic elliptic
functions of Jacobi and Weierstrass.

\section{The Einstein-Poisson Equation for Static  Gravitational Fields, the Active Gravitational Mass and the Fredholm Alternative} \label{sec3}

The central object in GR is the symmetric metric tensor $g_{\mu\nu}$ which plays a double r\^{o}le:
\begin{itemize}
\item[i)]
it determines the space-time geometry via the line element 
$
\ds^2=g_{\mu\nu}(x^\rho)\dx^\mu\dx^\nu$,
\item[ii)]
it represents the gravitational field, i.e. its 10 components replace the single scalar potential in Newton's gravity.
\end{itemize}
(Greek indices $\mu,\nu, \rho$ etc. run over 0,1,2,3, Latin indices i,j,k etc. go from 1 to 3.) 
The 4-vector
$
x^\mu=(x^0,x^1,x^2,x^3):=(ct,x,y,z)
$
labels an event along a world line of an observer at time $t$ and at the spatial point $\vec{x}=(x,y,z)$. $c$ is the speed of light in vacuo. Our sign convention for the Lorentzian metric is $(+---)\cdot g_{\mu \alpha}g^{\alpha \nu}=\delta^\nu_\mu:=1$ for $\mu=\nu$ and zero otherwise).
Given the {\em energy-momentum} (stress-energy) {\em tensor} $T_{\mu\nu}$ for matter, e.g. for a star, $g_{\mu\nu}(x^\rho)$ has to be determined as a solution of {\em Einstein's field equations}
\begin{equation} \label{equ3.1}
R_{\mu\nu}-\frac{1}{2} g_{\mu\nu}R-\Lambda g_{\mu\nu}=\frac{8 \pi G}{c^4}T_{\mu\nu} \ .
\end{equation}
(Here $R_{\mu\nu}:=R^\alpha_{\mu\alpha\nu}$ denotes the Ricci tensor, $R^\alpha_{\mu \beta \nu}$ the Riemann tensor (defined in terms of the Christoffel symbols), and $R:=g^{\mu\nu}R_{\mu\nu}$ is the Ricci scalar. $G$ is Newton's gravitational constant, and $\Lambda$ denotes Einstein's cosmological constant, see below.) For a perfect relativistic fluid $T_{\mu\nu}$ is given by
\begin{equation} \label{equ3.2}
T_{\mu\nu}=(\varepsilon + p) u_\mu u_\nu/c^2-pg_{\mu\nu} \ ,
\end{equation}
where $\varepsilon = \rho c^2$ is the energy (resp. mass) density and $p$ the pressure of the fluid. $u^\mu:=c \dx^\mu / \ds$ is the four-velocity $(u_\mu u^\mu=c^2)$.\\

The cosmological constant $\Lambda$ can be considered as a fundamental constant with dimension ${\rm(length)}^{-2}$ whose actual value cannot be reliably calculated at present but may be calculable in a future theory of quantum gravity. If the term $-\Lambda g_{\mu\nu}$ in (\ref{equ3.1}) is brought to the right-hand side, one can define an energy-momentum tensor $T^\Lambda_{\mu\nu}$ by
\begin{equation} \label{equ3.3}
\Lambda g_{\mu\nu}=:\frac{8\pi G}{c^4} T^\Lambda_{\mu\nu}=\frac{8\pi G}{c^4} \left[ (\varepsilon_\Lambda+p_\Lambda)u_\mu u_\nu / c^2-p_\Lambda g_{\mu\nu}\ \right] \ ,
\end{equation}
which for $\Lambda >0$ is in modern cosmology interpreted as ''dark energy'' possessing positive energy density $\varepsilon_\Lambda$ and {\em negative} pressure $p_\Lambda$ defined by
\begin{equation} \label{equ3.4}
\varepsilon_\Lambda:= \frac{\Lambda c^4}{8\pi G} \ , \ p_\Lambda:=-\varepsilon_\Lambda \ .
\end{equation}
In the widely accepted standard model of cosmology, the so-called $\Lambda$CDM-model, dark energy with the properties (\ref{equ3.4}) amounts to $ \approx$ 68\% of the total energy and is responsable for the observed  accelerated expansion of the universe \cite{11}.
The remaining energy of the universe consists of radiation (photons and neutrinos), baryonic and cold-dark matter such that the effective total energy momentum tensor $T^{\tot}_{\mu\nu}$ is still given by (\ref{equ3.2}) but expressed in terms of the total energy density 
$\varepsilon_{\tot}$ and the total pressure $p_{\tot}$.\\

Here our main concern is not cosmology but rather the gravitational field of a non-rotating, electrically neutral spherical body like a star or a black hole in a toroidal universe which can be described by one or several perfect fluids in terms of a total energy momentum tensor $T_{\mu\nu}^{\tot}$. The latter will include $T^\Lambda_{\mu\nu}$ in the case $\Lambda \neq 0$. From the Einstein equations (\ref{equ3.1}) one deduces $R=-\frac{8\pi G}{c^4}T^{\tot}$ with $T^{\tot}:=\left(T^{\tot}\right)^\mu_\mu$ which allows us to write these equations in the equivalent form
\begin{equation} \label{equ3.5}
R_{\mu\nu}=\frac{8\pi G}{c^4} \left(T^{\tot}_{\mu\nu}-\frac{1}{2}g_{\mu\nu}T^{\tot}\right) \ .
\end{equation}

Let us consider now a system for which one can find a coordinate system such that all components of $g_{\mu\nu}$ are independent of  time. Strictly speaking, this is only possible for the gravitational field produced by a single body. For example, in the case of a pair of black holes we know that the mutual gravitational attraction can lead to a coalescence of two black holes, i.e. their orbital inspiral and merger, and subsequent final black hole ringdown accompanied by gravitational-wave emission \cite{7}.
There are, however, important applications where a ''test-body'' 
 (for example a planet) moves in the field of a very massive body
 (for example the sun) and the metric can be considered in a very good approximation as being time-independent. With $g_{ij}=:-\gamma_{ij}$ we then obtain for the line element
\begin{equation} \label{equ3.6}
\ds^2=c^2 g_{00}(\vec{x})\dt^2+2cg_{0i}(\vec{x})\dt\dx^i-\gamma_{ij} (\vec{x})\dx^i\dx^j \ .
\end{equation}
If $g_{0i}(\vec{x})\neq 0$, this metric is called {\em stationary} which is, for example, the case for a rotating star. Such systems are not invariant under time reversal, $t \to - t$. Below we shall consider {\em static} space-times which are time-reversal invariant and thus one must have $g_{0i}\equiv 0$.\\

For the  stationary metric (\ref{equ3.6}), the Christoffel symbols and thus $R_{\mu\nu}$ simplify considerably,  and the Einstein eqs. (\ref{equ3.5}) can be conveniently written as three groups corresponding to $R^0_0,R^i_0$ and $R^i_j$.
Here we give only the $R^0_0$-equation (see \cite{21} for the remaining eqs.)
\begin{equation} \label{equ3.7}
\frac{1}{\sqrt{g_{00}}} \Delta_{LB} \sqrt{g_{00}}+ \frac{g_{00}}{4} f_{ij}f^{ij}=\frac{8\pi G}{c^4}\Bigl(\frac{\varepsilon_{\tot}+p_{\tot}}{1-\beta^2} - \frac{\varepsilon_{\tot}-p_{\tot}}{2} \Bigr) \ .
\end{equation}
$\Delta_{LB}$ denotes the 3-dimensional {\em Laplace-Beltrami operator} with respect to the spatial metric $\gamma_{ij}$ and is given by $(\gamma:=\det(\gamma_{ij}))$
\begin{equation} \label{equ3.8}
\Delta_{LB}:=\frac{1}{\sqrt{\gamma}} \frac{\partial}{\partial x^j}\Bigl(\sqrt{\gamma}\gamma^{ji}\frac{\partial}{\partial x^i}\Bigr) \ ,
\end{equation}
and the antisymmetric 3-tensor $f_{ij}$ is defined as
$$
f_{ij}:=\frac{\partial}{\partial x^i}\Bigl( \frac{g_{0j}}{g_{00}}\Bigr)-\frac{\partial}{\partial x^j} \Bigl(\frac{g_{0i}}{g_{00}}\Bigr) \ .
$$
Furthermore $\beta^2=|\vec{v}|^2/c^2$, where $|\vec{v}|^2$ is the square of  the 3-dimensional velocity in a space with the spatial metric $\gamma_{ij}$, i.e. $|\vec{v}|^2=v_iv^i=\gamma_{ij}v^jv^i, v^i=c(\dx^i/\ds)$.\\

Let us now consider a star or black hole at rest for which the 3-velocity $\vec{v}$ of matter is zero, the 4-velocity is given by $u^\mu=\delta^\mu_0 / \sqrt{g_{00}}$ and the space-time is {\em static}, i.e.
\begin{equation} \label{equ3.9}
\ds^2=c^2g_{00}(\vec{x})\dt^2-\gamma_{ij}(\vec{x})\dx^i\dx^j \ .
\end{equation}
Then the tensor  $f_{ij}$ is identically zero and one obtains from (\ref{equ3.7}), putting $\beta \equiv 0$, the {\em Einstein-Poisson (EP) equation}
\begin{equation} \label{equ3.10}
\frac{1}{\sqrt{g_{00}}} \Delta_{LB}\sqrt{g_{00}}=\frac{4\pi G}{c^4} (\varepsilon_{\tot}+3p_{\tot}) \ .
\end{equation}
Here the following remarks are in order:

\noindent
i)
 Eq. (\ref{equ3.10}) is in Einstein's theory of  gravitation the relativistic analogue of the Poisson equation for Newton's gravity, and therefore we call  it  Einstein-Poisson equation. It holds for static gravitational fields which are characterized by the existence of a time-like Killing vector field that is orthogonal to the hypersurface $t$=const. (For an alternative invariant derivation and interesting historical remarks,  see \cite{22}.) The close analogy with Newton's theory becomes also evident if one considers the energy $E$ and the acceleration of a point particle (planet) of mass $m$ at rest. From $E=mc^2 g_{0\mu} u^\mu$ one infers that
\begin{equation} \label{equ3.11}
\Phi(\vec{x}):=\frac{c^2}{2}(g_{00}(\vec{x})-1)
\end{equation}
plays the r\^{o}le of the static {\em relativistic potential} resp. $V:=m\Phi$ is the {\em static potential energy} at $\vec{x}$ of the planet since
\begin{equation} \label{equ3.12}
E=mc^2 \sqrt{g_{00}}=mc^2 \sqrt{1+\frac{2}{c^2}\Phi} =mc^2+V(\vec{x})+\cdots \ ,
\end{equation}
where $mc^2$ is the rest energy of the planet. Furthermore, the geodesic equation gives for the {\em force} $\vec{F}$ acting on the planet at rest $(d\tau:=\sqrt{g_{00}} \dt =$ proper time element of the planet, $\Gamma^i_{\mu\nu}$= Christoffel symbol)
$$
F^i=m \frac{d^2x^i}{d\tau^2}= - m \Gamma^i_{\mu\nu}  \frac{\dx^\mu}{d\tau} \frac{\dx^\nu}{d\tau} \quad (i=1,2,3) \ ,
$$
i.e.
\begin{equation} \label{equ3.13}
\vec{F}= - \frac{mc^2}{\sqrt{g_{00}}} \vec{\nabla}\sqrt{g_{00}}= - \frac{mc^2}{2g_{00}} \vec{\nabla} g_{00}= - \frac{m}{g_{00}} \vec{\nabla}\Phi= - \vec{\nabla}V+\cdots \  .
\end{equation}

\noindent
ii)
An important difference between Einstein's and Newton's theory of gravitation, which turns out to be crucial in  a torus universe, concerns the source of gravity \cite{22, 23}. In Newton's theory,  the total mass resp. the total mass density $\rho_{\tot}(\vec{x})$ is the source of gravitational fields. In a relativistic theory this is replaced by the total energy resp. the total energy density $\varepsilon_{\tot}=\rho_{\tot}c^2$.  But in Einstein's field equations (\ref{equ3.1}) the source of gravity is the energy-momentum tensor $T_{\mu\nu}$ which in the case of perfect fluids is given  by (\ref{equ3.2}).
Thus the total pressure $p_{\tot}$ contributes together with the total energy density in the combination $\varepsilon_{\tot}+3 p_{\tot}$
(see eq. (\ref{equ3.10})),  as pointed out first by Levi-Civit\'a \cite{24} and first studied by Tolman \cite{25}.
The latter showed that the total energy $E_{\tot}$ of a sphere of perfect fluids in  a state of stable or metastable equilibrium  is given  by
\begin{equation} \label{equ3.14}
E_{\tot}=\int (\varepsilon_{\tot}+3 p_{\tot}) \sqrt{g_{00}} \sqrt{\gamma} d^3 x \ ,
\end{equation}
where $\sqrt{\gamma}d^3 x$ is the proper 3-dimensional volume element (see eqs. (\ref{equ3.8}) and (\ref{equ3.9})).
The additional factor $\sqrt{g_{00}}$ in the above integral is in agreement with the EP-equation (\ref{equ3.10}) after multiplication by this factor.  Furthermore, Tolman obtained from (\ref{equ3.14}) an approximate expression for the energy of a sphere of perfect fluids in a weak gravitational field (i.e.~$\Phi (\vec{x})$ defined in (\ref{equ3.11}) being small) consisting of two parts, the first being the total proper energy of the material out of which the sphere is composed, and the second the well-known Newtonian expression $\frac{1}{2} \int \rho \Phi d^3 x$ for its gravitational potential energy.

Since the source of gravity is in Einstein's theory given by the specific combination $\hat{\rho}_{\tot}:=\rho_{\tot}+3p_{\tot}/c^2$, it is called the {\em active gravitational mass density} \cite{26}.\\

\noindent
iii) There are only a few exact solutions known to the EP eq. (\ref{equ3.10}) (for a given active gravitational mass density 
$\hat{\rho}_{\tot}$ and the associated spatial metric $\gamma_{ij}$). The most famous solutions are the static, spherically symmetric Schwarzschild \cite{1} and Schwarzschild-(anti)de Sitter \cite{2, 3} metrics for a black hole with $\Lambda=0$ resp. $\Lambda \neq 0$, and the stationary, axially symmetric Kerr metric \cite{4} for a rotating black hole (see also e.g. \cite{27}). To the best of our knowledge, there is no exact solution known to (\ref{equ3.10}) for a general toroidal universe. In the next section, we therefore introduce an iterative (perturbative) approach which will give us a well-defined approximate solution for weak gravitational fields in terms of triply-periodic functions. The existence and uniqueness of  the solutions satisfying periodic b.c. is guaranteed by a theorem known as the Fredholm alternative.

Consider on the fundamental domain $\Omega$ the {\em Poisson equation}
\begin{equation} \label{equ3.15}
\Delta_{LB}u(\vec{x})=S (\vec{x}) \ ,
\end{equation}
where the ''source'' $S(\vec{x})$ is a given triply-periodic function (see (\ref{equ2.6})). Here we assume that the operator 
$\Delta_{LB}$ defined in (\ref{equ3.8}) acts on the Hilbert space $H:=L^2(\Omega)$ with the inner product 
$$
\langle f,g \rangle:=\int\limits_\Omega \bar{f}g\sqrt{\gamma}d^3 x \  (f,g \in L^2(\Omega)  ,\mbox{ triply-periodic)} \ .
$$
 Then $-\Delta_{LB}$ is a positive self-adjoint elliptic operator whose spectrum is purely discrete since $\Omega$ is compact, $-\Delta_{LB}u_n=\lambda_n u_n$. We seek functions $u(\vec{x})\in L^2(\Omega)$ which are solutions of (\ref{equ3.15}) and satisfy the periodic b.c. (\ref{equ2.6}). Then the following {\em Fredholm alternative} \cite{28,29} holds which may also be called ''Fredholm dichotomy'' \cite{30}:
\begin{itemize}
\item[a)]
either $\Delta_{LB}u_0=0$ has only a trivial solution, then eq.~(\ref{equ3.15}) has a uniquely defined solution $u(\vec{x})$ for any given $S(\vec{x})$,
\item[b)]
or $\Delta_{LB}u_0(\vec{x})=0$ has a nontrivial solution, then the Poisson eq.~(\ref{equ3.15}) has a solution if, and only if, the  source $S(\vec{x})$ is orthogonal to $u_0(\vec{x})$, i.e. we have the {\em integrability condition}
\begin{equation} \label{equ3.16}
\langle S, u_0 \rangle = 0 \ .
\end{equation}
The solution $u(\vec{x})$ is unique up to an additive function $c_0u_0(\vec{x})$, where $c_0$ is an arbitrary constant.\\
\end{itemize}

Applied to our case, it is known that the eigenspaces are finite dimensional and that there exists exactly one isolated eigenvalue $\lambda_0=0$ whose eigenfunction is constant $(\parallel u_0 \parallel = 1)$
\begin{equation} \label{equ3.17}
u_0=|\Omega|_0^{-1/2} \mbox{  with  } |\Omega|_0:= \int\limits_\Omega \sqrt{\gamma} d^3 x \ .
\end{equation}
It therefore follows that the Poisson equation (\ref{equ3.15}) has a solution iff the source satisfies (\ref{equ3.16}), i.e. the {\em integrability condition}
\begin{equation} \label{equ3.18}
\int\limits_\Omega S(\vec{x})\sqrt{\gamma(\vec{x})} d^3 x = 0 \ .
\end{equation}
The solution is unique up to an arbitrary additive constant.  Let us add that the condition (\ref{equ3.16}) resp. (\ref{equ3.18}) can also be derived if one expands $S(\vec{x})$ and $u(\vec{x})$ in a Fourier series with respect to the complete orthonormal basis $u_n(\vec{x})$ of $H$.\\

That the existence of the zero mode is obstructive to finding a solution to (\ref{equ3.15}) becomes clear if one tries to construct the Green's function for $\Delta_{LB}$ under the prescribed b.c..
Since the Green's function of $\Delta_{LB}$ would be defined as the integral kernel of the inverse operator, but the inverse of $\Delta_{LB}$ does not exist due to the eigenvalue $\lambda_0=0$, the usual construction fails, and no Green's function exists in $H:=L^2(\Omega)$. It is possible, however, to define a {\em Green's function in the generalized sense} \cite{29}.
Consider the decomposition $H=\ker(-\Delta_{LB})\oplus H^\perp$, where $H^\perp:= (\ker (-\Delta_{LB}))^\perp$ is the orthogonal complement in $H$ of the zero mode eigenspace. Then the restriction of $-\Delta_{LB}$ to $H^\perp$ is in this space a self-adjoint strictly positive operator and thus its inverse exists in $H^\perp$.  The Green's function in the generalized sense, $K(\vec{x},\vec{y})$, has to satisfy the periodic b.c. and the differential equation
\begin{equation} \label{equ3.19}
\Delta_{LB,\vec{x}} K(\vec{x},\vec{y})=(\gamma (\vec{x})\gamma(\vec{y}))^{-1/4} \delta_{\T^3}(\vec{x}-\vec{y})-\frac{1}{|\Omega |_0} \ ,
\end{equation}
where $\delta_{\T^3}(\vec{x}-\vec{y})$ is the periodic delta function on $\T^3$ (see (\ref{equ4.6})).
The  rhs of (\ref{equ3.19}) is in accordance with the completeness relation in $H^\perp$ (in the sense of distributions)
$$
\sum\limits_{\lambda_n > 0} u_n(\vec{x})\bar{u}_n (\vec{y})=\sum\limits_{\lambda_n \ge 0} u_n(\vec{x})\bar{u}_n(\vec{y}) -
u_0(\vec{x})\bar{u}_0(\vec{y}) \ ,
$$
where in our case  the zero-mode term is given by the constant $u_0(\vec{x})\bar{u}_0(\vec{y})$ $
\equiv u_0^2=1/| \Omega|_0$. To single out a particular function $K$, one can require \cite{29} $\langle K,u_0\rangle=0$, i.e. for $u_0=$ const.
\begin{equation} \label{equ3.20}
\int\limits_\Omega K(\vec{x}, \vec{y})\sqrt{\gamma(\vec{y})}d^3y=0 \ .
\end{equation}
Since the solution $u(\vec{x})$ of (\ref{equ3.15}) is determind only up to an arbitrary constant, it can be fixed by means of the condition $\langle u,u_0\rangle = 0$, i.e.
\begin{equation} \label{equ3.21}
\int\limits_\Omega u(\vec{x})\sqrt{\gamma}d^3x=0 \ .
\end{equation}
Then  the unique solution of the Poisson equation (\ref{equ3.15}) is given by
\begin{equation} \label{equ3.22}
u(\vec{x})=\int\limits_\Omega K(\vec{x}, \vec{y})S(\vec{y})\sqrt{\gamma(\vec{y})}d^3 y \ ,
\end{equation}
iff the integrability condition (\ref{equ3.18}) is satisfied. As shown, the condition (\ref{equ3.18}) is required for mathematical consistency. But (\ref{equ3.18}) has also a  very deep physical interpretation as an {\em equilibrium} or {\em stability condition} (see below and section 4).\\

Coming back to the EP eq. (\ref{equ3.10}), it is important to note that this equation is {\bf not} of the inhomogeneous type as the Poisson eq. (\ref{equ3.15}) but rather takes the following form of a homogeneous equation
\begin{equation} \label{equ3.23}
\Delta_{LB}\sqrt{g_{00}(\vec{x})} = \frac{4\pi G}{c^4} (\varepsilon_{\tot}(\vec{x})+3p_{\tot}(\vec{x}))\sqrt{g_{00}(\vec{x})} \ .
\end{equation}
Nevertheless, the generalized Green's function $K(\vec{x},\vec{y})$ defined by (\ref{equ3.19}) can still be used to write
(\ref{equ3.23}) in the equivalent form of the following {\em Fredholm integral equation}
\begin{equation} \label{equ3.24}
\sqrt{g_{00}(\vec{x})} = 1+\frac{4\pi G}{c^4} \int\limits_{\Omega}K(\vec{x},\vec{y}) (\varepsilon_{\tot}(\vec{y})
+ 3p_{\tot}(\vec{y})) 
 \sqrt{g_{00}(\vec{y})} \sqrt{\gamma(\vec{y})} d^3 y \ ,
\end{equation}
iff the {\em integrability condition}
\begin{equation} \label{equ3.25}
\int\limits_{\Omega}(\varepsilon_{\tot}+3p_{\tot})\sqrt{g_{00}}\sqrt{\gamma}d^3x=0
\end{equation}
is satisfied. Comparing (\ref{equ3.25}) with Tolman's formula (\ref{equ3.14}), we see that the integrability condition states that the total energy $E_{\tot}$ (expressed in terms of the active gravitational energy density and the gravitational field $\sqrt{g_{00}}$) of a spherical star of perfect fluid being in a state of stable equilibrium has to be identically zero.\\

The integral equation (\ref{equ3.24}) can be solved in the limit of weak gravitational fields ($g_{00}\approx 1$, see 
eq.~(\ref{equ3.11})) by iteration which leads for the dominant term to the first-order EP equation (\ref{equ4.3}) discussed in the next section which now has the form of a Poisson equation as (\ref{equ3.15})
but for the ordinary Laplacian $\Delta$. The convergence of the Neumann series in terms of the iterated kernels is not studied here (see the expansion (\ref{equ4.2})).

\section{The Static Gravitational Field of a Star in a Flat Torus Universe in the Weak Field Limit}
\label{sec4}

A general static space-time metric in a flat toroidal universe can in cartesian coordinates always be written in the following form (see (\ref{equ3.9}) with $\gamma_{ij}\sim \delta_{ij}$ and (\ref{equ3.11}))
\begin{equation} \label{equ4.1}
\ds^2 = c^2\Bigl(1+\frac{2}{c^2}\Phi (\vec{x}) \Bigr)\dt^2 - \Bigl(1-\frac{2}{c^2}\Psi (\vec{x})\Bigr)
\Bigl( \dx^2+\dy^2+\dz^2\Bigr) \ .
\end{equation}
$\Phi$ and $\Psi$ depend on the dimensional parameters $c,G$ and $M$ ($M$ denotes the mass of a star), the ''effective size'' $L$ of the torus universe of the order of the Hubble length $L_H \approx 1.4 \times 10^{26}$ $m$ and the space vector $\vec{x}=(x,y,z)$. From these quantities we define the Schwarzschild radius (of the star with mass $M$) $r_s:=2GM/c^2$ and the dimensionless ''coupling strength'' $\kappa:=r_s/L$. Furthermore, we introduce the dimensionless coordinate vector $\vec{\xi}:=\vec{x}/L$. We then expand $\Phi$ and $\Psi$ in a power series in $\kappa$ (''perturbation series'') as
\begin{equation} \label{equ4.2}
\frac{2}{c^2}\Phi = \sum\limits^\infty_{n=1}\Phi_n (\vec{\xi})\kappa^n
\end{equation}
(and analogously for $\frac{2}{c^2}\Psi$ with $\Psi_1 \equiv \Phi_1$). The {\em weak field limit} is then defined by considering only the first terms $\Phi_1\kappa$ and $\Psi_1 \kappa$, with $\kappa \ll |\vec{\xi}| \ll 1$. To get an idea of the order of magnitude: 
\begin{itemize}
\item[i)]
for the earth with mass $M_\oplus \approx 5.97 \times 10^{24}$ kg one obtains $r_s \approx 8.87\times 10^{-3}$ m and $\kappa\approx 6.5 \times 10^{-29}$ (using $L\equiv L_H$),
\item[ii)]
for the sun with mass $M_\odot\approx 1.99 \times 10^{30}$ kg, $r_s=2.95\times 10^3$ m, $\kappa\approx 2.2 \times 10^{-23}$,
\item[iii)]
for the supermassive black hole $Sgr A^\star$ at the centre of the milky way with mass $M\approx 4.4 \times 10^6 M_\odot, r_s \approx 1.3 \times 10^{10}$ m, $\kappa\approx 10^{-16}$.
\end{itemize}
It is seen that typical $\kappa$-values are extremely small and thus the first term in (\ref{equ4.2}) should be a good approximation.
Denoting this term by $\frac{2}{c^2}\Phi_{\T^3}(\vec{x})$, our task is to find a solution to the corresponding {\em first-order EP equation} (reintroducing $\vec{x}$ as variable)
\begin{equation} \label{equ4.3}
\Delta \Phi_{\T^3}(\vec{x})=4\pi G (\rho_M(\vec{x})+\hat{\rho}_{X}(\vec{x})) \ ,
\end{equation}
where $\Delta$ denotes the euclidean Laplacian $\Delta:=\frac{\partial^2}{\partial x^2}+\frac{\partial^2}{\partial y^2}+\frac{\partial^2}{\partial z^2}$.
Here we have assumed that the spherical star is well described by a perfect fluid with mass density $\rho_M(\vec{x})$, pressure $p_M \equiv 0$, and we have kept the {\em active gravitational mass density}  
$\hat{\rho}_{X}  :=(\varepsilon_{X}+3p_{X})/c^2$ of an additional perfect fluid which is of topological origin and has to be determined from the integrability condition  (\ref{equ3.18}). Considering the EP equation (\ref{equ4.3}) on the covering space $\E^3, \rho_M(\vec{x})$ and $\hat{\rho}_{X}(\vec{x})$ are triply-periodic functions, and thus we have to determine a triply-periodic function $\Phi_{\T^3}(\vec{x})$ solving eq.~(\ref{equ4.3}).\\

To find the desired solution, let us make the following ansatz (based on the method of images) in the form of a Poincar\'e series (see eqs. (\ref{equ2.3}), (\ref{equ2.5}) and (\ref{equ2.6}))
\begin{equation} \label{equ4.4}
\Phi_{\T^3} (\vec{x}):=\sum\limits_{\gamma_{\vec{n}} \varepsilon\Gamma} f \left( \gamma_{\vec{n}} (\vec{x} )\right)=\sum\limits_{\vec{n}} f\left( \vec{x}-\vec{\rho}(\vec{n}) \right) \  ,
\end{equation}
where $f(\vec{x})$ is a suitable function on $\E^3$. (Throughout this paper, the sum $\sum\limits_{\vec{n}}$ denotes a triple sum over all $\vec{n}=(n_1,n_2,n_3)\in \Z^3$, a prime being inserted, $\sum_{\vec{n}}{}^{' } $, when the term $n_1=n_2=n_3=0$ has to be omitted from  the summation.)
One is tempted to make the choice $f(\vec{x}):=\Phi_N(\vec{x}):= - GM/r \quad (r=|\vec{x}|)$, where $\Phi_N(\vec{x})$ denotes Newton's gravitational  potential, which leads to the ansatz 
\begin{equation} \label{equ4.5}
\sum\limits_{\vec{n}} \Phi_N ( \vec{x}-\vec{\rho}(\vec{n}))   =-\frac{GM}{r}-GM\sum_{\vec{n}}{}^{' }  \frac{1}{|\vec{x}-\vec{\rho}(\vec{n})|} \ .
\end{equation}
Very informally, this ansatz appears to solve eq. (\ref{equ4.3}) (using $\Delta(\frac{1}{r})=- 4\pi\delta^3(\vec{x}))$ corresponding to $\hat{\rho}_{X}\equiv 0$ and
\begin{equation} \label{equ4.6}
\begin{array}{lll}
4\pi G \rho^{\T^{3}}_M(\vec{x})\!\!&:=&\!\!4\pi GM\sum\limits_{\vec{n}}\delta^3(\vec{x}-\vec{\rho}(\vec{n}))\\ 
\!\!&\equiv&\!\!4\pi GM \delta_{\T^3}(\vec{x}) \ ,
\end{array}
\end{equation}
where $\delta_{\T^3}(\vec{x})$ denotes the periodization of the 3-dimensional delta function.
$\rho^{\T^3}_M(\vec{x})$ can be interpreted as the mass density of a star with mass $M$ considered as a pointlike particle.
Unfortunately, the triple sum (\ref{equ4.5}) over infinitely many Newton potentials is divergent which can be seen  from the {\em Epstein zeta function} (see  eq.~(\ref{equ4.40})) $\sum_{\vec{n}}{}^{' }  |\vec{n}|^{-s}, s \varepsilon \C$, which converges absolutely in the half plane $\Re s > 3$, but is divergent for $\Re s \le 3$. (Actually, this function has an analytic continuation as  a meromorphic function to the whole complex plane with a single pole  at $s=3$.) Due to this divergence, the ansatz (\ref{equ4.5}) is useless for a computation of $\Phi_{\T^3}(\vec{x})$ and the application of the Laplacian on (\ref{equ4.5}) acting on the individual terms is not allowed. But, more seriously, (\ref{equ4.5}) does not satisfy (\ref{equ3.18})!\\

There are, however, ways to regularize the sum (\ref{equ4.5}) in such a way that a well-defined solution to the first order EP equation is obtained in terms of absolutely convergent series. In  the following we shall present three different regularizations which give, however, the same unique solution (apart from an additive constant) in different representations.

\subsection{The Gravitational Field in Terms of the Appell Zeta Function} 

Our first regularization of the  ansatz (\ref{equ4.5}) is based on the work of Appell \cite{16, 17, 18} dating back to 1884!
From  the generating function of the Legendre polynomials $P_\ell(\cos \gamma)$ we obtain for $r<\rho$ the absolutely convergent expansion $(|P_\ell (\cos \gamma)|\le 1)$
\begin{eqnarray} \label{equ4.7}
\frac{1}{|\vec{x}-\vec{\rho}|}&=&\frac{1}{\Bigl[\vec{\rho}^2-2(\vec{x}\cdot \vec{\rho})+\vec{x}^2\Bigr]^{1/2}} \nonumber\\
\!&=& \!\frac{1}{\rho} \frac{1}{\Bigl[1-2\left(\frac{r}{\rho}\right)\cos \gamma +\left(\frac{r}{\rho}\right)^2 \Bigr]^{1/2}}  \\
\!&=&\! \frac{1}{\rho} \sum\limits^\infty_{\ell = 0} P_\ell (\cos \gamma) \Bigl(\frac{r}{\rho}\Bigr)^\ell \ ,  \nonumber
\end{eqnarray}
where $\rho=|\vec{\rho}(\vec{n})|$ (see eqs. (\ref{equ2.3}), (\ref{equ2.4})) and $\gamma=\gamma(\vec{x},\vec{n})\in [0,2\pi]$ denotes the angle between $\vec{x}$ and $\vec{\rho}(\vec{n})$. Using for $\vec{x}$ and $\vec{\rho}$ spherical coordinates
$(r, \vartheta,\varphi)$ resp. $(\rho,\vartheta^\prime,\varphi^\prime)\equiv (\rho (\vec{n}),\vartheta^\prime (\vec{n}), \varphi^\prime (\vec{n}))$, we have
\begin{eqnarray} \label{equ4.8}
\cos \gamma\!&=&\! \frac{x(\vec{b_1}\cdot \vec{n})+y(\vec{b}_2 \cdot \vec{n})+z(\vec{b_3}\cdot \vec{n})}{r\rho}\\
\!&=&\! \cos \vartheta \cos \vartheta^\prime + \sin \vartheta \sin \vartheta^\prime \cos (\varphi-\varphi^\prime) \ .  \nonumber
\end{eqnarray}
Since the sum $\sum_{\vec{n}}{}^{' }  (\rho (\vec{n}))^{-s}$ converges absolutely for $\Re s > 3$ (see eq.~(\ref{equ4.40})), we can regularize the triple sum in (\ref{equ4.5}) by subtracting the first three terms of the expansion (\ref{equ4.7}).
We then define {\em Appell's $Z$-function} \cite{16} by
\begin{equation} \label{equ4.9}
Z_A(\vec{x}):=\frac{1}{r}+\sum_{\vec{n}}{}^{' }  \Bigl[\frac{1}{|\vec{x}-\vec{\rho}(\vec{n})|}-\frac{1}{\rho(\vec{n})}-
 P_1(\cos\gamma)\Bigl(\frac{r}{\rho^2}\Bigr)-P_2(\cos \gamma)\Bigl(\frac{r^2}{\rho^3}\Bigr)\Bigr] \ .
\end{equation}

Here the triple sum is absolutely convergent for all $\vec{x}\in\R^3$ except at the origin and at the congruent points $\vec{\rho}(\vec{n})$.
$Z_A(\vec{x})$ possesses the following properties:
\begin{itemize}
\item[i)]
$Z_A(-\vec{x})=Z_A(\vec{x})$,
\item[ii)]
$Z_A(\vec{x})$ is regular for $\vec{x}\in\R^3$ except at the origin resp. at the points $\vec{\rho}(\vec{n})$ where it has poles $\frac{1}{r}$ resp. $\frac{1}{|\vec{x}-\vec{\rho}(\vec{n})|}$ with residue~1. Thus $Z_A(\vec{x})$ has in every PEP of the lattice $\cL$ exactly one singularity,
\item[iii)]
$Z_A(\vec{x})$ is, however, {\bf not} triply-periodic, since one can show \cite{16} that all second derivatives 
$\left(\frac{\partial^2}{\partial x^2},\frac{\partial^2}{\partial x \partial y} \ \mbox{ etc.} \right)$ of the difference $Z_A(\vec{x}+\vec{a}_k)-Z_A(\vec{x}) \ (k=1,2,3$; see eq.~(\ref{equ2.2})) are identically zero and thus this difference can only be a linear function in $\vec{x}$
\end{itemize}
\begin{equation} \label{equ4.10}
Z_A(\vec{x}+\vec{a}_k)-Z_A(\vec{x}) = B_k+\vec{x}\cdot \vec{C}_k \quad (k=1,2,3) \  .
\end{equation}
Here the 12 constants $B_k$ and $C_{k\ell} \ (k,\ell=1,2,3)$ can be obtained from the absolutely convergent series expansion (\ref{equ4.9}) of $Z_A$:
\begin{equation} \label{equ4.11}
\left. \begin{array}{l} 
\vec{C}_k=\left(C_{k1},C_{k2},C_{k3}\right):=2\vec{\nabla}Z_A\left(\frac{\vec{a}_k}{2}\right) \\
 B_k:=\frac{1}{2} \left(\vec{a}_k\cdot \vec{C}_k \right) \  .
\end{array}\right\}(k=1,2,3)
\end{equation}
In addition one has the consistency relations $\vec{a}_1\cdot \vec{C}_2=\vec{a}_2\cdot\vec{C}_1$ and cycl.\\

In oder to construct from $Z_A(\vec{x})$ a  triply-periodic function, let us consider a real symmetric $(3\times 3)$-matrix $D$ and the associated quadratic form $D[\vec{x}]:=\sum\limits^3_{k,\ell=1}x_kD_{k\ell} x_\ell$. Then Appell \cite{16} has shown that the matrix $D$ can be determined in such a way that $D[\vec{x}]$ satisfies for all $\vec{x}\in \R^3$ the relation (see eq. (\ref{equ4.10}))
\begin{equation} \label{equ4.12}
D[\vec{x}+\vec{a}_k]-D[\vec{x}]=-B_k-\vec{x}\cdot \vec{C}_k \quad(k=1,2,3) \ . 
\end{equation}
A Taylor expansion gives
\begin{equation} \label{equ4.13}
\vec{\nabla}D[\vec{a}_k]= - \vec{C}_k, \ D[\vec{a}_k]=-B_k \ (k=1,2,3) \ .
\end{equation}
Here the first nine linear equations are compatible due to the relations mentioned below eq. (\ref{equ4.11}) and allow to determine the matrix $D$.\\

With the help of the quadratic form $D[\vec{x}]$, {\em Appell's zeta function} is defined by \cite{18}
\begin{equation} \label{equ4.14}
\zeta_A(\vec{x}):=Z_A(\vec{x})+D[\vec{x}] \ .
\end{equation}
$\zeta_A$ has the following properties:
\begin{itemize}
\item[i)]
$\zeta_A(-\vec{x})=\zeta_A(\vec{x})$,
\item[ii)]
$\zeta_A(\vec{x})$ is {\em triply-periodic}
\begin{equation} \label{equ4.15}
\zeta_A(\vec{x}+\vec{a}_k)=\zeta_A(\vec{x}) \ (k=1,2,3) \ .
\end{equation}
\item[iii)]
$\zeta_A(\vec{x})$ is regular for $\vec{x}\in \R^3$ except at the origin and at the congruent points $\vec{\rho}(\vec{n})$.
\item[iv)]
$\zeta_A(\vec{x})$ satisfies the {\em Poisson equation}
\end{itemize}
\begin{equation} \label{equ4.16}
\Delta \zeta_A(\vec{x})=-4\pi \Bigl(\delta_{\T^3}(\vec{x})-\frac{1}{|\Omega |}\Bigr)
\end{equation}
and thus satisfies {\bf automatically} the integrability condition (\ref{equ3.18}) 
(adopted to the weak-field limit, i.e. $\gamma\equiv 1$)
according to the Fredholm alternative discussed in Sect.~3.\\

\noindent
{\em Proof:} (Appell \cite{18}).
The definition (\ref{equ4.14}) gives $\Delta \zeta_A=\Delta Z_A+\Delta D=- 4\pi \delta_{\T^3}(\vec{x})+K$ with $K:=2 Tr D$. Without any  computation it follows that $K$ cannot vanish since Appell showed \cite{18} that a triply-periodic function $f(\vec{x})$ having in the fundamental domain only a single pole, cannot satisfy $\Delta f = 0$. In order to compute $K$, consider the fundamental domain $\Omega$ containing the origin and the ball $B(\hat{\varepsilon})$ with centre at the origin and having a very small radius $\hat{\varepsilon}$. Apply Green's theorem to the region $\Omega \setminus B$
$$
\int\limits_{\Omega \setminus B} \Delta \zeta_A d^3 x=\int\limits_{\partial \Omega} \frac{\partial \zeta_A}{\partial n} d\sigma + \int\limits_{\partial B}\frac{\partial \zeta_A}{\partial n} d\sigma \ .
$$
Then the volume integral on the $\ell h s$ is equal to $K\left(|\Omega|-\frac{4\pi}{3}\hat{\varepsilon}^3\right)$.
The surface integral over $\partial \Omega$ is zero since $\zeta_A$ is triply-periodic and the values of the outer normal derivatives $\partial \zeta_A / \partial n$ at congruent points on $\partial \Omega$ agree up to a sign. In the surface integral on $\partial B$ we insert $\zeta_A=\frac{1}{r}+ \eta_A$, where $\eta_A$ stays finite if $\hat{\varepsilon}$ goes to zero, and thus the integral tends to $4\pi$ in this limit and we obtain $K=4\pi / |\Omega|$.
\qed\\

We thus arrive at the desired solution,  which we denote by $\Phi_A$\\
(''A'' stands for Appell), of the first-order EP eq. (\ref{equ4.3}), \begin{equation} \label{equ4.17}
\Phi_A(\vec{x}):= - GM\zeta_A (\vec{x}) \ ,  
\end{equation}
which explicitly reads
\begin{eqnarray} \label{equ4.18}
\ \ \ \ \ \ \ \ \ \ \ \ \Phi_A(\vec{x})=\!\!&-&\!\! \frac{GM}{r}-GM\sum\limits^3_{k,\ell=1}x_kD_{k\ell}x_\ell\\
\!\!&-&\!\! GM \sum_{\vec{n}}{}^{' }  \Bigl[\frac{1}{|\vec{x}-\vec{\rho}(\vec{n})|}-\frac{1}{\rho}-\frac{1}{2}(3\cos^2 \gamma - 1) \frac{r^2}{\rho^3}\Bigr] \ . \nonumber
\end{eqnarray}
$\Phi_A$ describes the gravitational field of a non-rotating star in a static state of stable or metastable equilibrium in the far-field
limit such that the field is weak and the star can be treated as a point-like object (localized source), i.e. $\rho_M(\vec{x})=M\delta_{\T^3}(\vec{x})$. (Compare the Poisson eqs. (\ref{equ4.3}) and (\ref{equ4.16}).) This part of the gravitational source
in  (\ref{equ4.3}) is responsable for the first term in (\ref{equ4.18}) which is the Newtonian potential of a body of mass $M$ and thus {\bf defines} also in our case the constant $M$ to be the mass of the relativistic source.
This definition is the standard one in the case of an infinite universe for which space-time is asymptotically flat (see e.g. \cite{27}).
Furthermore, we have assumed that the material out of which the star is composed  is a perfect pressureless fluid (''dust''), i.e. $p_M=0$.\\

A novel and very remarkable property of the field $\Phi_A$ is that its source contains in addition to $\rho_M$ a second term, see (\ref{equ4.16}), which has (taking the definition (\ref{equ4.17}) into account) to be identified with the active gravitational mass  densitity $\hat{\rho}_X$ in (\ref{equ4.3}), i.e.
\begin{equation} \label{equ4.19}
\hat{\varepsilon}_X := \hat{\rho}_X c^2= \varepsilon_X + 3 p_X
\equiv - \frac{Mc^2}{|\Omega|}=: \varepsilon_{\ttop}+ 3 p_{\ttop} \ .
\end{equation}
Since this term is required by the integrability condition (\ref{equ3.18}) which in turn is related to the non-trivial topology of the  torus universe, we call this term a {\em topological} (''top'') {\em active gravitational energy density}.
Assuming that $\varepsilon_{\ttop} > 0$, the pressure $p_{\tot}$ must be {\bf negative} and thus this source acts like a sort of {\em ''topological dark energy''} in analogy to the dark energy in cosmology which is well described by Einstein's cosmological constant $\Lambda$ (see Sec.~3). We therefore define a (positive) {\em specific topological constant} $\Lambda_{\ttop}:=\frac{4\pi GM}{|\Omega|c^2}=\frac{2\pi r_s}{|\Omega|}$ and (see (\ref{equ3.4})) $\varepsilon_{\ttop}:=\frac{\Lambda_{\ttop}c^4}{8\pi G},p_{\ttop}:=-\varepsilon_{\ttop}$.  Note that $\Lambda_{\ttop}$ is not a universal constant of nature like Einstein's $\Lambda$, but rather depends on the mass $M$ of the star and the volume $|\Omega|$ of the torus universe.
With $L\approx L_H, |\Omega | \approx L^3_H$ (see the discussion after eq.~(\ref{equ4.1})),  one obtains $\Lambda_{\ttop}\approx 2\pi \kappa/ L^2_H \approx 10^{-52}\kappa m^{-2}$ which gives even for the Milky Way with $M\approx 10^{12} M_{\odot}$ the very small value $\Lambda_{\ttop}\approx 10^{-63}m^{-2}$ to be compared with the upper limit of Einstein's cosmological constant $\Lambda \le 10^{-52}m^{-2}$. Note, however, that $\Lambda_{\ttop}$ is much larger if one considers an early epoch where the universe has been much smaller. Finally, let us mention that the topological term (\ref{equ4.19}) has been obtained directly from Appell's triply-periodic zeta function without demanding the integrability condition (\ref{equ3.18}).  \\
        
Another important property of the triply-periodic field $\Phi_A(\vec{x})$ is its {\em anisotropy} inherited from the global non-isotropic geometry of the torus universe. This is best seen by rewriting the triple sum in (\ref{equ4.18}) as a multipole expansion. Inserting for $|\vec{x}-\vec{\rho}|^{-1}$ the Legendre expansion (\ref{equ4.7}), one obtains for the triple sum
$$
\sum\limits^\infty_{\ell=2} \sum_{\vec{n}}{}^{' }  P_{2\ell}(\cos \gamma) \frac{r^{2\ell}}{\rho^{2\ell+1}} \ ,
$$
where only the Legendre polynomials of even order contribute. (This is seen by replacing in the summation $\vec{n}$ by $-\vec{n}$ and using $\rho (-\vec{n})=\rho(\vec{n}), \cos \gamma (\vec{x},-\vec{n})= - \cos \gamma (\vec{x},\vec{n})$ (see (\ref{equ4.8})) and
$$
P_\ell (- \cos \gamma)=(-1)^\ell P_\ell (\cos \gamma) .)
$$
 Then we get with the help of  the addition theorem
$$
P_\ell (\cos \gamma(\vec{x},\vec{n}))=\frac{4\pi}{2\ell +1} \sum\limits^\ell_{m=-\ell} \bar{Y}_{\ell m} (\vartheta^\prime,\varphi^\prime) Y_{\ell m}(\vartheta,\varphi)
$$
for the spherical harmonics on $\s^2$ the {\em multipole expansion}
\begin{eqnarray} \label{equ4.20}
\Phi_A(\vec{x})=\!\! &-&\!\! \frac{GM}{r}-GM \sum\limits^3_{k,\ell=1} D_{k\ell}x_kx_\ell \\
\!\!&- &\!\! 4\pi GM \sum\limits^\infty_{\ell=2}\sum\limits^{2 \ell}_{m=- 2 \ell}
 \frac{S_{2 \ell,m}}{4 \ell+1} Y_{2 \ell, m} (\vartheta, \varphi)r^{2 \ell} \ , \nonumber
\end{eqnarray}
where the {\em multipole moments} are defind by  ($\ell \ge 2$; see also (\ref{equ2.4}))
\begin{equation} \label{equ4.21}
S_{2\ell, m}:= \sum_{\vec{n}}{}^{' }  \frac{\bar{Y}_{2\ell,m}(\vartheta^\prime (\vec{n}),\varphi^\prime (\vec{n}))}{(\rho(\vec{n}))^{2\ell + 1}}\\
= \sum_{\vec{n}}{}^{' }  \frac{\bar{Y}_{2\ell,m}(\vartheta^\prime (\vec{n}),\varphi^\prime (\vec{n}))}{(\hat{Q}[\vec{n}])^{\ell + 1/2} }  \ .
\end{equation}
The expansion (\ref{equ4.20}) will be useful if it is sufficient to take only a few multipoles into account in computing, for example, the geodesic motion of a planet in the gravitational field $\Phi_A(\vec{x})$, see (\ref{equ3.13}). Obviously, the trajectories are much more complicated than those for the motion in the Schwarzschild \cite{31} or Schwarzschild-de Sitter metric \cite{32}. \\

The subtle structure of the gravitational field $\Phi_A(\vec{x})$ (for example, its equipotential surfaces $\Phi_A(\vec{x})$ = const.) is encoded in the multipole moments $D_{k\ell}$ and $S_{2\ell,m}=(-1)^m \bar{S}_{2\ell,-m}$ which reflect the spatial shape of the torus universe, i.e. the fundamental parallelepiped $\Omega \subset \E^3$ defined by the matrix $A$ resp. the Gram matrix $\hat{Q}=AA^T$ (see Sec.~2). As an illustration, let us discuss the first correction of order $O(r^2)$ to the Newtonian potential which is given by the quadratic form associated with the matrix $D$ defined in (\ref{equ4.12}) and (\ref{equ4.13}).  (It is expected that this term gives the dominant contribution for $r_s \ll r \ll L_H$, since the remaining multipole sum is of order $(GM/L_H)(r/L_H)^4$.) It is important to note that this term can never be zero due to the general relation $Tr D= 2\pi / | \Omega |$ (see the proof of (\ref{equ4.16})). Consider as a simple example a 3-torus universe whose fundamental domain is a rectangular PEP (a cuboid) with different edge lengths $L_1,L_2,L_3$. Then $\vec{\rho}(\vec{n})=(n_1 L_1,n_2 L_2,n_3L_3)$, the matrix $D$ is diagonal with {\bf different} diagonal elements
\begin{eqnarray}\label{equ4.22}
D_{11}=-\frac{1}{L_1}\frac{\partial}{\partial x} Z_A\Bigl( \frac{L_1}{2},0,0\Bigr),\\
D_{22}=-\frac{1}{L_2}\frac{\partial}{\partial y}Z_A\Bigl(0,\frac{L_2}{2},0\Bigr)  , \nonumber\\ 
D_{33}=-\frac{1}{L_3}\frac{\partial}{\partial z}Z_A\Bigl(0,0, \frac{L_3}{2}\Bigr) , \nonumber
\end{eqnarray}
and  the gravitational field is given by
\begin{equation} \label{equ4.23}
\Phi_A(\vec{x})=-\frac{GM}{r}-GM(D_{11}x^2+D_{22}y^2+D_{33}z^2)+O(r^4)
\end{equation}
which has still a complicated non-isotropic form.\\

The simplest case is obtained if the fundamental domain of the torus universe is a cube with edge length $L=L_1=L_2=L_3$, for which the matrix $D$ takes the simple form $D=(\frac{2\pi}{3})L^{-3}  \one_3$. Then $g^A_{00}:=1+\frac{2}{c^2}\Phi_A$ (see (\ref{equ3.11}) and (\ref{equ4.11})) is given by
\begin{equation} \label{equ4.24}
g^A_{00}=1-\frac{r_s}{r}-\frac{\Lambda_{\ttop}}{3}r^2-4\pi \Bigl(\frac{r_s}{L}\Bigr)\sum\limits^\infty_{\ell=2}\sum\limits^{2\ell}_{m=-2\ell}
\frac{\hat{S}_{2\ell,m}}{4\ell+1} Y_{2\ell,m}(\vartheta,\varphi)\Bigl(\frac{r}{L}\Bigr)^{2\ell} \ .
\end{equation}
Here the topological constant is given by $\Lambda_{\ttop}=2\pi r_s/L^3$ and the multipole moment $\hat{S}_{2\ell,m}$ by the Dirichlet series $(\ell \ge 2)$
\begin{equation} \label{equ4.25}
\hat{S}_{2\ell,m}:=\sum_{\vec{n}}{}^{' }  \frac{\bar{Y}_{2\ell,m}(\vartheta^\prime(\vec{n}),\varphi^\prime(\vec{n}))}{|\vec{n}|^{2\ell+1}} \ .
\end{equation}
At this point it is appropriate to discuss the main question of this paper: {\em do black holes exist in a torus universe?} 
Although we cannot give a final answer to this question, we think that the solution (\ref{equ4.24}) gives support to an affirmative answer. First of all, (\ref{equ4.24}) is the {\bf exact}
solution to the first order EP equation with respect to the very small expansion parameter \mbox{$\kappa \! :=r_s/L$.} (Note that $\Lambda_{\ttop}r^2=2\pi \kappa  (r/L)^2$.) Thus the remaining terms of order $\kappa^n(n\ge 2)$ are expected to be extremely small for $r \ll L$ and therefore negligible. But most important is the observation that the first 3 terms of (\ref{equ4.24}) are identical to the exact spherically symmetric {\em Schwarzschild-de~Sitter metric} if Einstein's cosmological constant $\Lambda$ is replaced by $\Lambda_{\ttop}$. Since the multipole sum in (\ref{equ4.24}) is  bounded by 
$\kappa \zeta_E(5)\left(\frac{r}{L}\right)^4 \left[1-\left(\frac{r}{L}\right)^2\right]^{-1} \approx 10.38 \kappa \left(\frac{r}{L}\right)^4$ ($\zeta_E$ is defined below in (\ref{equ4.40})) it follows
that this term too is for $r \ll L$ extremely small  and thus we end up with the approximate solution
\begin{equation} \label{equ4.26}
\bar{g}_{00}^A:=1-\frac{r_s}{r}-\frac{\Lambda_{\ttop}}{3} r^2 \ .
\end{equation}
This solution depends on the two parameters $r_s$ and $\Lambda_{\ttop}$ resp. on the mass $M$ and the torus length $L$. To see whether this solution possesses one or more horizons, we have to determine the zeros of $\bar{g}_{00}^A$ which leads to the cubic equation $r^3+3pr+2q=0$ with $p:=-1/\Lambda_{\ttop},q:=\frac{3}{2}(r_s / \Lambda_{\ttop})$ and the associated discriminant $d:=p^3+q^2 = - \left( 1-\frac{3}{4}r^2_s \Lambda_{\ttop}\right)/ \Lambda^3_{\ttop}< 0$ for $\frac{3}{4}r^2_s \Lambda_{\ttop}=\frac{3\pi}{2} (r_s/L)^3 \ll 1$. We then obtain three real solutions, where one solution is negative and the two positive solutions correspond to the {\em black hole horizon} $r_{BH}$ and the {\em cosmological horizon} $r_c$ with $r_{BH}<r_m<r_c$ given by
\begin{equation}\label{equ4.27}
\left\{ \begin{array}{l}
r_{BH}:= \frac{2}{\sqrt{\Lambda_{\ttop}}} \cos \bigl( \beta+\frac{\pi}{3}\bigr) = r_s \bigl(1+\frac{2\pi}{3}\bigl(\frac{r_s}{L}\bigr)^3+0 \bigl(\bigl(\frac{r_s}{L}\bigr)^5\bigr)\underset{ (L \to  \infty)}{ \longrightarrow}r_s   \\
r_m:= \bigl(\frac{3}{2} \frac{r_s}{\Lambda_{\ttop}}\bigr)^{1/3}=\bigl(\frac{3}{4\pi}\bigr)^{1/3} L\\
r_c:=  \frac{2}{\sqrt{\Lambda_{\ttop}}} \cos \bigl( \beta-\frac{\pi}{3}\bigr) =\sqrt{\frac{3}{2\pi}\frac{L}{r_s}} L-\frac{r_s}{2}+0 \bigl(\bigl(\frac{r_s}{L}\bigr)^{3/2}r_s\bigr) \underset{ (L\to \infty)}{\longrightarrow} \infty \ .
\end{array} \right.
\end{equation}
Here $r_m$ denotes the radius at which $\bar{g}_{00}^A$ has a relative maximum and the angle $\beta$ is defined as $\beta:=\frac{1}{3} \arccos \left(\frac{3}{2}r_s \sqrt{\Lambda_{\ttop}} \right)$.
Thus it holds $\bar{g}_{00}^A > 0$ for $r_{BH} <r<r_c$.
The approximation  (\ref{equ4.26}) is meaningful for $R \ll r \ll r_m$, where $R \gg r_{BH}$ is the radius of the compact object.

\subsection{The Gravitational Field in Terms of the Epstein Zeta Function}

Our second regularization of the Poincar\'e series (\ref{equ4.4}) resp. of the ansatz (\ref{equ4.5}) makes use of the Epstein zeta function \cite{19, 20}. The idea is to regularize the sum (\ref{equ4.5}) by replacing $|\vec{x}-\vec{\rho}|^{-1}$ by the complex power 
$|\vec{x}-\vec{\rho}|^{-s}, s \in \C, \Re s > 3$, instead of subtracting the divergent terms as done in (\ref{equ4.9}) following Appell \cite{16, 17, 18}. To keep the discussion simple, we consider the special case of a 3-torus whose fundamental domain is a cube of edge length $L$ and thus $\vec{\rho}=L\vec{n}$ resp. $|\vec{x}-\vec{\rho}|^{-1}=L^{-1} |\vec{\xi}-\vec{n}|^{-1}$ with $\vec{\xi}:=\vec{x}/L$. {\em Epstein's zeta function} is defined by (''E'' stands for Epstein)
\begin{equation}\label{equ4.28}
\zeta_E(s,\vec{\xi}):=\sum\limits_{\vec{n}} \frac{1}{\bigl|\vec{n}+\vec{\xi}\bigr|^{s}} \ , \ Re s > 3 \ .
\end{equation}
For $\vec{\xi} \not\in \Z^3$, the above Dirichlet series converges absolutely in the half-plane $\Re s > 3$ and uniformly in any compact subset. Following Riemann's method by which he derived the analytic continuation of his zeta function $\zeta(s)$, Epstein \cite{19} showed that (\ref{equ4.28}) has an analytic continuation into the whole complex {\em s}-plane as a meromorphic function possessing only a single pole at $s=3$ with residue $4\pi$. (Actually, Epstein defined $\zeta_E(s,\vec{\xi})$ for general dimension $D\in\N$, i.e. for $\vec{\xi} \in \R^D$, and proved the analytic continuation, in which case there exists a single pole at $s=D$ with residue $2\pi^{D/2}/\Gamma \left(\frac{D}{2}\right)$.) Thus the analytic continuation of $\zeta_E$ is regular at $s=1$, and we therefore {\bf define} the gravitational field $\Phi_E(\vec{x})$ by
\begin{equation}\label{equ4.29}
\Phi_E (\vec{x}):=-\frac{GM}{L} \zeta_E(1,\vec{\xi}) \ .
\end{equation}
To obtain the analytic continuation, (\ref{equ4.28}) is rewritten as the Mellin transform of a generalized $\Theta$-function (see eq. (\ref{equ4.36})), then the Mellin integral is split into $\int^1_0 \dt + \int^\infty_1 \dt$ and it is observed that the second integral is absolutely convergent in the whole {\em s}-plane for $\vec{\xi} \not\in \Z^3$ and thus is an entire function of $s$. In the first integral one inserts a general $\Theta$-transformation formula (see eq.~(\ref{equ4.37}), derived by the Poisson summation formula) and then makes the variable substitution $t\rightarrow 1/t$. As a result, the first integral produces the simple pole at $s=3$, and one is left with an integral which is also absolutely convergent in the whole $s$-plane. One thus arrives at the following {\em integral representation for Epstein's zeta function} valid in the whole $s$-plane \cite{19}
\begin{eqnarray}\label{equ4.30}
\ \ \ \ \ \ \ \pi^{-s/2}\Gamma\Bigl(\frac{s}{2}\Bigr) \zeta_E(s,\vec{\xi})\!\!&=&\!\! \frac{2}{s-3}+\sum\limits_{\vec{n}} \Gamma^*\Bigl(\frac{s}{2} , \pi |\vec{n}-\vec{\xi}|^2   \Bigr)\\
\!\!&+&\!\! \sum_{\vec{n}}{}^{' }  \cos (2\pi (\vec{n}\cdot \vec{\xi})) \Gamma^* \Bigl(\frac{3-s}{2},\pi \vec{n}^2 \Bigr) \ . \nonumber
\end{eqnarray}
Here $\Gamma^*(s,x)$ denotes the modified incomplete gamma function
\begin{equation}\label{equ4.31}
\Gamma^*(s,x):=\int\limits^\infty_1 \dt \ t^{s-1} e^{-xt}=x^{-s}\Gamma (s,x) \ ,
\end{equation}
where $\Gamma(s,x)$ is  the standard incomplete gamma function \cite{33}. With
$\Gamma^* \left( \frac{1}{2},x\right) = \sqrt{\pi} \frac{\erfc(\sqrt{x})}{\sqrt{x}}$ and $\Gamma^*(1,x)=\frac{e^{-x}}{x}$ we obtain from (\ref{equ4.29}) and (\ref{equ4.30}) for the  solution of  the first-order EP eq.~(\ref{equ4.3}) the series expansion
\begin{eqnarray}\label{equ4.32}
\Phi_E(\vec{x})=\frac{GM}{L}\!\!&-&\!\!\frac{GM}{L} \sum\limits_{\vec{n}} \frac{\erfc \bigl( \sqrt{\pi} |\vec{\xi}-\vec{n}|\bigr)}{|\vec{\xi}-\vec{n}|}\\
\!\!& -&\!\! \frac{GM}{\pi L} \sum_{\vec{n}}{}^{'} \cos(2\pi (\vec{n}\cdot\vec{\xi}))
\frac{e^{-\pi \vec{n}^2}}{\vec{n}^2} \  . \nonumber
\end{eqnarray}
$\Phi_E(\vec{x})$ has the following properties:
\begin{itemize}
\item[i)]
$\Phi_E(-\vec{x})=\Phi_E (\vec{x})$,
\item[ii)]
it is triply-periodic,
\item[iii)]
the triple sums over $\vec{n}$ converge absolutely for $\vec{\xi} \not\in \Z^3$,
\item[iv)]
the first  sum has a pole at $r=0$ and all conjugate points with residue $-GM$,
\item[v)]
the solution (\ref{equ4.32}) is unique up to an arbitrary additive constant and thus it must be identical to the solution
$\Phi_A(\vec{x})$, eqs. (\ref{equ4.18}) resp. (\ref{equ4.20}) (derived from Appell's zeta function), up to an additive constant,
\item[vi)]
$\Phi_E(\vec{x})$ satisfies the Poisson equation and therefore fulfills automatically the integrability condition
\end{itemize}
\begin{equation}\label{equ4.33}
\Delta \Phi_E(\vec{x})=4\pi GM \delta_{\T^3} (\vec{x})- \frac{4\pi GM}{L^3} \  .
\end{equation}

\noindent
{\em Proof:} i) and ii) are obvious, also iii) is obvious for the second sum in (\ref{equ4.32}). For the first sum we recall the  asymptotic expansion for  the completed error function \cite{33}
$$
\frac{\erfc (\sqrt{\pi}x)}{x}=\frac{e^{-\pi x^2}}{\pi x^2} \Bigl[  1-\frac{1}{2\pi x^2} + O\Bigl(\frac{1}{x^4} \Bigr) \Bigr]  \  , \  x \to \infty \ .
$$
iv) To see the pole structure, we split off the term with $\vec{n}=\vec{0}$ in the first sum in (\ref{equ4.32}), use the identity $\erfc (x)=1-\mbox{erf} (x)$ and the series expansion of the standard error function \cite{33}
$$
\mbox{erf} (x) = \frac{2}{\sqrt{\pi}} \sum\limits^\infty_{m=0} \frac{(-1)^m}{(2m+1)m!} x^{2m+1}
$$
to obtain
\begin{eqnarray} \label{equ4.34}
\ \ \ \ \ \ \ \ \ \ \Phi_E(\vec{x})\!&=&\! \frac{3 GM}{L}-\frac{GM}{r}+\frac{2GM}{L}\sum\limits^\infty_{m=1} \frac{(-1)^{m}\pi^m}{(2m+1)m!} \Bigl(\frac{r}{L} \Bigr)^{2m}\\
\!&-&\! \frac{GM}{L}  \sum_{\vec{n}}{}^{' } \Bigl[ \frac{\erfc (\sqrt{\pi}|\vec{\xi}-\vec{n}|)}{|\vec{\xi}-\vec{n}|} + \frac{\cos(2\pi (\vec{n}\cdot\vec{\xi}))}{\pi}  \frac{e^{-\pi \vec{n}^2}}{\vec{n}^2}\Bigr]  \  .  \nonumber
\end{eqnarray}
From  this one obtains immediately $g_{00}^E:=1+\frac{2}{c^2}\Phi_E$ which for $r\ll L$ differs up to order $(r/L)^2$ from the approximate solution $\bar{g}_{00}^A$, eq. (\ref{equ4.26}),  only by the extremely small additive term $3\kappa$.\\

Let us mention that the solution $\Phi_E(\vec{x})$ is ideally suited for a numerical evaluation since the triple sums in  (\ref{equ4.34}) converge very rapidly due to the exponential factors $e^{-\pi|\vec{n}-\vec{\xi}|^2}$ resp. $e^{-\pi|\vec{n}|^2}$. On the other hand, it is quite cumbersome to obtain from (\ref{equ4.34}) the multipole structure analytically which in the case of $\Phi_A$ was straightforward.\\

From (\ref{equ4.34}) one also easily derives that $g_{00}^E$ approaches for $\vec{x}$ fixed in the limit of an infinite torus volume, i.e. for $L\to \infty$, the Schwarzschild metric $g_{00}^s:=1-r_s/r$.\\

To derive the Poisson equation (\ref{equ4.33}), we require the following relations: $\Delta(1/r)=-4\pi\delta^3(\vec{x})$ and with 
$
\Delta=L^{-2}\Delta_{\vec{\xi}}=L^{-2} \bigl( \frac{\partial^2}{\partial \xi_1^2}+\frac{\partial^2}{\partial \xi_2^2}+\frac{\partial^2}{\partial \xi_3^2}\bigr): $
\begin{eqnarray*}
\Delta_{\vec{\xi}} \biggl( \frac{\erf (\sqrt{\pi}|\vec{\xi}|)}{|\vec{\xi}|}\biggr) \!\!&=&\! \!-\, 4\pi e^{-\pi |\vec{\xi}|^2}\\
\Delta_{\vec{\xi}}  \biggl( \frac{\erfc (\sqrt{\pi}|\vec{\xi}-\vec{n}|)}{|\vec{\xi}-\vec{n}|}  \biggr)
\!\!& = & \!\!+\, 4 \pi  e^{-\pi |\vec{\xi}-\vec{n}|^2}\\
\Delta_{\vec{\xi}} \bigl(e^{2\pi i(\vec{n}\cdot\vec{\xi})} \bigr)\! \!&=&\!\! -\, 4\pi^2 \vec{n}^2  e^{2\pi i (\vec{n}\cdot\vec{\xi})} \  .
\end{eqnarray*}
Here we used the fact that in the last sum in (\ref{equ4.34}) $\cos (2\pi (\vec{n}\cdot\vec{\xi}))$ can be replaced by 
$e^{2\pi i (\vec{n}\cdot\vec{\xi})}$. We then obtain (we consider $\vec{\xi}$ in the fundamental domain $\Omega$)
\begin{equation}\label{equ4.35}
\begin{array}{lllll}
\Delta \Phi_E \! \!&=&\! \!4 \pi GM \delta^3(\vec{x})  - \frac{4\pi GM}{L^3}  \bigl(e^{-\pi |\vec{\xi}|^2} 
 + \sum\limits_{\vec{n}}{}^{' }  e^{-\pi |\vec{\xi}-\vec{n}|^2}  - \sum\limits_{\vec{n}}{}^{' }e^{2\pi i(\vec{n}\cdot \vec{\xi} )-\pi \vec{n}^2 }  \bigr)   \\
\! \!&=& \!\! 4\pi GM \delta^3(\vec{x})-\frac{4\pi GM}{L^3} \bigl( \Theta(1; \vec{\xi},\vec{0})- \Theta(1;\vec{0},\vec{\xi})+1 \bigr) \ ,
\end{array}
\end{equation}
where we have introduced the {\em generalized $\Theta$-function} $(\vec{\xi},\vec{h} \in \R^3, t>0)$
\begin{equation} \label{equ4.36}
\Theta \bigl(t; \vec{\xi},\vec{h}\bigr):=\sum\limits_{\vec{n}} e^{-\pi|\vec{n}-\vec{\xi}|^2 t+2\pi i (\vec{n}\cdot\vec{h})} \ ,
\end{equation}
 which appeared in the Mellin transform of $\zeta_E$. From Poisson summation one obtains the {\em transformation formula} \cite{19}
\begin{equation} \label{equ4.37}
\Theta \bigl(t; \vec{\xi},\vec{h}\bigr)=\frac{e^{-2\pi i\bigl(\vec{\xi} \cdot \vec{h}\bigr)}}{t^{3/2}} \Theta \Bigl(\frac{1}{t}; \vec{h},-\vec{\xi}\Bigr)
\end{equation}
 which yields for $t=1, \vec{h}=\vec{0}$
 $$
 \Theta \Bigl(1; \vec{\xi},\vec{0}\Bigr)=\Theta \Bigl(1; \vec{0},-\vec{\xi}\Bigr)=\Theta \Bigl(1;\vec{0},\vec{\xi}\Bigr)
 $$
 and thus the $\Theta$-terms in (\ref{equ4.35}) cancel out and we obtain the Poisson eq. (\ref{equ4.33}) in the fundamental domain. Since $\Phi_E$ is triply-periodic, this implies also the general form (\ref{equ4.33}).

\subsection{ The Gravitational Field in Terms of the Generalized Green's Function}
Our third method to solve the first-order EP equation (\ref{equ4.3}) is based on a regularization of the generalized Green's function. As an example, we consider again a cubic torus universe with volume $|\Omega|=L^3$. It is easy to see that the {\em eigenfunctions} of $-\Delta$ on $\T^3, u_{\vec{n}}(\vec{x}):=L^{-3/2} e^{2\pi i(\vec{n} \cdot \vec{x})/L},
\vec{n}\in \Z^3$, with the {\em discrete eigenvalues} $\lambda_{\vec{n}}:=(2\pi/L)^2 \vec{n}^2$ form  a complete orthonormal basis of $L^2(\Omega)$. Since there is a zero eigenvalue, we have to consider, as discussed in Sec.~3, the restriction of $-\Delta$ to the Hilbert space $H^\perp = (\ker(-\Delta))^\perp$. In order to guarantee absolute convergence, we define the following {\em regularization of the generalized Green's function} $(s \in \C, \Re s >3)$
\begin{equation} \label{equ4.38}
K_s(\vec{x},\vec{y}):=- \sum_{\vec{n}}{}^{' } 
\frac{u_{\vec{n}}(\vec{x})\overline{u_{\vec{n}}(\vec{y})}}{\lambda_{\vec{n}}^{s/2}}
 = - \frac{L^{s-3}}{(2\pi)^s} \zeta_E \Bigr(s,\vec{0},\frac{\vec{x}-\vec{y}}{L}\Bigr) \ .
\end{equation}
Here we have introduced another special case of {\em Epstein's zeta function} \cite{19} $(s\in\C, \Re s > 3, \vec{h} \in \R^3)$
\begin{equation} \label{equ4.39}
\zeta_E(s,\vec{0},\vec{h}):= \sum_{\vec{n}}{}^{' } \frac{e^{2\pi i (\vec{n}\cdot \vec{h})}}{|\vec{n}|^s} \ .
\end{equation}
The notation indicates that the general Epstein zeta function \cite{19} $\zeta_E(s,\vec{g},\vec{h})$ depends on $s$ and the ''characteristic'' $| {\vec{g} \atop \vec{h}}|$ with $\vec{g},\vec{h} \in \R^3$. Thus the zeta function mentioned after eq. 
(\ref{equ4.6}) corresponds to $| {\vec{0}\atop\vec{0}}|$, i.e.
\begin{equation} \label{equ4.40}
\zeta_E(s):=\zeta_E(s,\vec{0},\vec{0}):=\sum_{\vec{n}}{}^{' } \frac{1}{|\vec{n}|^s} \quad (\Re s > 3) \ ,
\end{equation}
whereas the Epstein zeta function $\zeta_E(s,\vec{\xi})$ defined in (\ref{equ4.28}) corresponds to the characteristic 
$| {\vec{\xi}\atop\vec{0}}|$.\\

Note that the kernel $K_s(\vec{x},\vec{y})$ is for $s=4,6,\cdots$ identical to the well-known ''iterated kernels'' \cite{29} which converge absolutely and uniformly in both $\vec{x}$ and $\vec{y}\in \R^3$. For $s=2$, eq. (\ref{equ4.38}) takes the form of Mercer's theorem which, however, cannot be applied here as it could in the case of ordinary differential equations \cite{29}.\\

Following again Riemann's method, as discussed after eq. (\ref{equ4.28}), Epstein proved \cite{19} that $\zeta_E(s,\vec{0},\vec{h})$ has a representation analogous to (\ref{equ4.30}) in terms of $\Gamma^\star(s,x)$ valid in  the whole $s$-plane from which one concludes that it is an entire function of $s$. Thus we  are led to identify the {\em gravitational field} $\Phi_G(\vec{x})$ (''G'' refers to Green's function), i.e. the solution of   (\ref{equ4.3}), with $4\pi GMK_2(\vec{x},\vec{0})$, 
which in terms of the analytic continuation of the Epstein zeta function (\ref{equ4.39}), see (\ref{equ4.38}), is then {\bf defined } by
\begin{equation} \label{equ4.41}
\Phi_G(\vec{x}):= - \frac{GM}{\pi L} \zeta_E\left(2,\vec{0},\frac{\vec{x}}{L}\right) \ .
\end{equation}
Epstein also proved \cite{19} that his zeta functions satisfy a {\em functional equation} (analogous to the famous functional equation for the Riemann zeta function $\zeta(s)$) which in our case is given by
\begin{equation} \label{equ4.42}
\pi^{-s/2}\Gamma \Bigl(\frac{s}{2}\Bigr)\zeta_E\Bigl(s,\vec{0},\vec{\xi}\Bigr)=\pi^{-\frac{3-s}{2}}\Gamma\Bigl( \frac{3-s}{2}\Bigr)\zeta_E\Bigl(3-s,\vec{\xi}\Bigr) \ .
\end{equation}
For $s=2$ this gives $\frac{1}{\pi}\zeta_E(2,\vec{0},\vec{\xi})=\zeta_E(1,\vec{\xi})$ and we thus obtain from 
 (\ref{equ4.41}) and  (\ref{equ4.29})
\begin{equation} \label{equ4.43} 
\Phi_G(\vec{x})\equiv \Phi_E(\vec{x}) \ .
\end{equation}
Due to the orthogonality of the eigenfunctions  $u_{\vec{n}}(\vec{x})$, we obtain from (\ref{equ4.38}) $\int\limits_\Omega
K_s (\vec{x},\vec{y})d^3y=0$. This implies the weak field normalization (see (\ref{equ3.20}) and (\ref{equ3.21}) for $\gamma=\det(\gamma_{ij})\to 1$)
\begin{equation} \label{equ4.44}
\int\limits_\Omega \Phi_G(\vec{x})d^3x=0 \ ,
\end{equation}
which fixes the arbitrary additive constant in $\Phi_G$ and $\Phi_E$. A comparison with $\Phi_A$ in (\ref{equ4.18}) yields for the cubic torus
\begin{equation} \label{equ4.45}
\Phi_E(\vec{x})=\Phi_A(\vec{x})+\Phi_0 \ ,
\end{equation}
where the constant field $\Phi_0$ is given by
\begin{eqnarray}\label{equ4.46}
\ \ \ \  \ \ \ \ \Phi_0\!\!&:=&\!\!-\frac{GM}{L}\zeta_E(1)=\frac{GM}{L}\Bigl[ 3-\sum_{\vec{n}}{}^{' }\Bigl(\frac{\erfc (\sqrt{\pi}|\vec{n}|)}{|\vec{n}|}+\frac{e^{-\pi\vec{n}^2}}{\pi\vec{n}^2} \Bigr)\Bigr]\\
\!\!&=&\!\!\frac{GM}{L} \Bigl[ 3-\sum\limits^\infty_{m=1}r_3(m)\Bigl(\frac{\erfc (\sqrt{\pi m})}{\sqrt{m}} +\frac{e^{-\pi m}}{\pi  m}\Bigr)\Bigr]
\nonumber \\
\!\!&=&\!\! \frac{GM}{L} \cdot 2.8372974 \ . \nonumber
\end{eqnarray}
Here $r_3(m)$ denotes the number of representations of $m\in \N$ as a sum of 3 squares of integers, where representations with different orders and different signs are counted as distinct. Obviously, $r_3(m)$ is the {\em multiplicity} of the {\em distinct eigenvalues} $\bar{\lambda}_m:=(2\pi/L)^2m, 0 = \bar{\lambda}_0 < \bar{\lambda}_1 < \bar{\lambda}_2 < \cdots (r_3(0)=1)$ of $-\Delta$ on the cubic torus. For example, $r_3(1)=6,r_3(2)=12,r_3(3)=8,r_3(4)=6,r_3(5)=24$.

The number theoretical function $r_3(m)$ is a very irregular function which has already been studied by Gau{\ss}. Defining the {\em counting function} $N(\lambda):= \# \{\lambda_{\vec{n}} \le \lambda\}$ for the eigenvalues $\lambda_{\vec{n}}$ of the cubic torus, one obtains {\em Weyl's law} (see e.g. \cite{34})
\begin{equation}\label{equ4.47}
N(\lambda)=\sum\limits_{\bar{\lambda}_m \le \lambda} r_3(m)= \frac{|\Omega |}{6\pi^2} \lambda^{3/2}+ 0(\sqrt{\lambda}),\lambda \to \infty \ .
\end{equation}
This follows from the small-$t$ asymptotics of $\Theta(t; \vec{0},\vec{0})$ given in (\ref{equ4.37}) using the Tauberian theorem of Karamata, since $\Theta(\frac{4\pi}{L^2} t; \vec{0},\vec{0})$ is identical to the trace of the heat kernel of $-\Delta$ on the cubic torus with volume $|\Omega |=L^3$. The same $t$-asymptotics  determines also the pole of $\zeta_E(s)$ at $s=3$, since $\zeta_E(s)$ is the Mellin transform of $\Theta(t; \vec{0},\vec{0})$ as shown by Epstein \cite{19} using Riemann's method. Due to the exponential convergence of the lattice sums in (\ref{equ4.46}), we have computed the numerical value given in (\ref{equ4.46}) by neglecting the terms with $m\ge 6$ using the Tables given in \cite{35}. (The same Tables have also been used for the value of $\zeta_E(5)$ cited before (\ref{equ4.26}).)

\section{Concluding Remarks}
We presented a perturbative approach for the static gravitational field of a spherical non-rotating star of mass $M$ in a universe whose space form is a general flat 3-torus defined by its volume $|\Omega|$ (resp. an effective length scale $L$) and 5 Teichm\"uller parameters. The perturbative expansion is in terms of a power series in the parameter $\kappa:=r_s/L \ (r_s$ = Schwarzschild radius) which is even for supermassive black holes extremely small $(\kappa \le 10^{-13}$ for $L \approx L_H$ = Hubble length). Thus the first-order term should provide a very good approximation in the far-field limit, 
$r_s \ll R<r \ll L$, where $R$ is the radius of the star. Since the well-known ansatz (\ref{equ4.4}) in form of a Poincar\'e series (method of images) is in general divergent (see the discussion after eq.~(\ref{equ4.5})) resp. not absolutely convergent (see eq.~(\ref{equ4.39})), we used 3 different methods to regularize the triple sums over $\vec{n}\in\Z^3$ which are based on Appell's triply-periodic zeta function \cite{16,17,18} resp. on the analytic continuation of Epstein's triply-periodic zeta function \cite{19, 20}. Appell's zeta function is not very well-known, and we therefore described its construction and main properties in some detail following Appell's original papers using, however, a more compact vector formulation introduced in Sec.~2. By contrast, Epstein's zeta function is well-known and thus we only sketched its derivation. In this way we obtained the two different representations $\Phi_A$, eqs. (\ref{equ4.18}) and (\ref{equ4.20}), and $\Phi_E$, eq. (\ref{equ4.32}), for the exact solution of the first-order Einstein-Poisson equation which both involve absolutely convergent series only. ($\Phi_A$ holds for the most general 3-torus, whereas $\Phi_E$ is given for simplicity for a cubic torus.) According to the Fredholm alternative discussed in Sec.~3, the solution is unique up to an additive constant field $\Phi_0$ which is confirmed by explicit calculation, see eqs. (\ref{equ4.45}, \ref{equ4.46}).\\

The first-order gravitational field in a finite toroidal universe possesses two remarkable properties which distinguishes it greatly from the standard spherically symmetric Schwarzschild (S) resp. Schwarzschild-de Sitter (S-dS) field in the infinite universe with $\M^3=\E^3$:\\

i) Let us consider the exterior region $(r>R)$ of a star. Then the standard black hole solutions in $\E^3$ are either solutions of the Einstein vacuum equations $R_{\mu\nu}=0$ (in the case of the S-metric) or of $R_{\mu\nu}=-\Lambda g_{\mu\nu}~ (\Lambda > 0$ in the case of the S-dS-metric) where Einstein's cosmological constant $\Lambda$ can take any value, i.e. it is {\bf not} determined by GR. As discussed in Sec.~3, $\Lambda$ is then interpreted as determining dark energy with energy density $\varepsilon_\Lambda=\Lambda c^4/(8\pi G)$ and negative pressure $p_\Lambda = - \varepsilon_\Lambda$.\\

This is completely different in a torus universe. Here the vacuum equations $R_{\mu\nu}=0$ have {\bf no} non-trivial solutions at all! Instead the {\em Einstein equations} are 
\begin{equation} \label{equ5.1}
R_{\mu\nu}=-\Lambda_{\ttop}g_{\mu\nu} \ ,
\end{equation}
where we have defined  the positive constant
\begin{equation} \label{equ5.2}
\Lambda_{\ttop}:=2\pi \frac{r_s}{|\Omega|} \ ,
\end{equation}
which is completely fixed by the Schwarzschild radius $r_s$ (i.e. the mass $M$ of the star) and the spatial volume 
$|\Omega |$ of the torus universe. Since this is a direct and  unavoidable consequence of the non-trivial topology of 
$\T^3$, we call $\Lambda_{\ttop}$ the {\em topological cosmological constant} of a given star of mass $M$. The physical interpretation is this: a stable or metastable equilibrium of a star in a toroidal universe requires the existence of a sort of {\em topological dark energy} which fills the whole universe and possesses a negative active gravitational energy density $\varepsilon_{\ttop}+3p_{\ttop}$
 with positive energy density $\varepsilon_{\ttop} =\Lambda_{\ttop}c^4/(8\pi G)$ and negative pressure $p_{\ttop}=- \varepsilon_{\ttop}$.
In Sect.~3 we discussed in detail the connection between the physical equilibrium condition and the mathematically required integrability condition according to the Fredholm alternative. It is worthwhile to mention that we did not use these conditions in the derivation of the gravitational field since both the Appell and the Epstein zeta function fulfil these conditions automatically (see eqs. (\ref{equ4.16}) and (\ref{equ4.33})).\\

ii) Another important property of the exterior field derived here is its {\em anisotropy} reflecting the breaking of global rotational symmetries which is explicitly seen in the {\em multipole expansion} (\ref{equ4.20}). Here the {\em monopole} is equal to Newton's potential, there is no {\em dipole} contribution, and the first non-vanishing (and for $R\ll r \ll L$ dominant) term is the {\em quadrupole field} $\Phi_Q(\vec{x})$, say.
Defining the {\em quadrupole tensor}
\begin{equation} \label{equ5.3}
\hat{D}_{k\ell}:=D_{k\ell}-\frac{2\pi}{3| \Omega |} \delta_{k\ell} \ ,
\end{equation}
which is traceless due to the general relation $TrD=\frac{2\pi}{|\Omega |}$, we rewrite $\Phi_Q(\vec{x})$ as follows
\begin{equation} \label{equ5.4}
\Phi_Q(\vec{x}):= - \frac{\Lambda_{\ttop}c^2}{6}r^2-GM \sum\limits^3_{k,\ell=1}\hat{D}_{k\ell}x_kx_\ell \ .
\end{equation}
This shows that there is for all tori, independent of their particular shape, a non-vanishing spherically symmetric quadrupole field which depends  solely on $\Lambda_{\ttop}$ and agrees for $\Lambda_{\ttop}\to \Lambda$ with the corresponding term of the S-dS black hole solution \cite{2, 3}.
In addition, there is in general an {\em anisotropic part} due to the tensor $\hat{D}$ which vanishes in special cases, for example for a cubic torus, see (\ref{equ4.24}).\\

Let us consider the force $\vec{F}$ which a planet of mass $m$ at rest experiences in the gravitational field of a star of mass $M \gg m$. In the weak field limit, the force is given by (\ref{equ3.13}), and thus the monopole leads to the well-known attractive Newtonian force. The interesting result comes from the quadrupole field (\ref{equ5.4}), where the first term leads to a {\em repulsive} central force $(\Lambda_{\ttop}c^2/3)r\hat{e}_r$ which depends only on $\Lambda_{\ttop}$! Even more interesting is the force governed by the quadrupole tensor $\hat{D}$ which   is in general {\em anisotropic}.\\

At this point it should be mentioned that a weak acceleration effect due to multiple connectedness of the universe was heuristically discussed in \cite{36} for different topologies. In the case of the 3-torus, the calculations were carried out using an expression for the force derived from a divergent series analogous to (\ref{equ4.5}) which is plagued by serious convergence problems as discussed after eq. (\ref{equ4.5}). The authors claimed that in the case of a cubic 3-torus there is {\bf no} force proportional to $r$. This contradicts our result derived from a well-defined regularization. The authors  also discussed the topological space $\M^3=\R^2 \times \s^1$, for which there exists an exact Schwarzschild-like solution to Einstein's equations \cite{37}, and derived the expected {\em topological acceleration}.\\

Since we have no exact solution to the full non-linear Einstein equations which describes the space-time metric of black holes in a torus universe, the question asked in the title of this paper remains open. Nevertheless, based on the exact first-order solution we {\em conjecture} that black holes exist in spaces with the topology of 3-tori. Furthermore, we expect that the gravitational field of these black holes is in the far-field limit well approximated by the first-order solution and thus possesses a very similar anisotropic pattern described by a multipole expansion.
An observation of the anisotropy, for example by a measurement of the quadrupole, would be an important signature of a finite universe resp. of non-trivial topology because it would indicate a violation of {\em Birkhoff's theorem} \cite{38, 39}.
This theorem, which holds under certain conditions in the standard case $\M^3=\E^3$, states that the Schwarzschild solution \cite{1} is the unique static, spherically symmetric exterior solution for an isolated spherical star, and that the same holds for the Schwarzschild-de~Sitter metric if $\Lambda > 0$ \cite{2, 3}.
Closely related to this uniqueness property is the so-called {\em ''no-hair theorem''} \cite{40} which states that a Schwarzschild black hole is completely characterized by only one free parameter, the mass $M$, which is not considered as ''hair''. Similarly a Kerr black hole \cite{4} is fully determined by the mass $M$ and the angular momentum $\cJ$.
 In case black holes exist in a torus universe according to our conjecture, they could be considered as ''hairy black holes'' possessing 6 hair, namely they would be fully characterized by $M$ (not counted), the  volume $|\Omega |$ and the 5 Teichm\"uller parameters of the 3-torus. (For black holes in quantum gravity, see \cite{41}.)

\setlength{\parindent}{0mm}


\begin{thebibliography}{XX}

\bibitem[1]{1}
K.  Schwarzschild:
\emph{\"Uber das Gravitationsfeld eines Massenpunktes nach der Einsteinschen Theorie}.
Sitzungsber. Kgl. Preu{\ss}. Akad. Wiss. zu Berlin, phys.-math. Klasse 1916, 189--196.


\bibitem[2]{2}
F.  Kottler:
\emph{\"Uber die physikalischen Grundlagen der Einsteinschen Gravitationstheorie}.
Ann. Phys. (Berlin) \textbf{56} (1918), 401--461.

\bibitem[3]{3}
H. Weyl:
\emph{\"Uber die statischen kugelsymmetrischen L\"osungen von Einsteins ''kosmologischen Gravitationsgleichungen''}.
Phys. Z. \textbf{20} (1919), 31--34.

\bibitem[4]{4}
R.P. Kerr:~\emph{Gravitational field of a spinning mass as an example of algebraically special metrics}.
Phys. Rev. Lett. \textbf{11} (1963), 237--238.

\bibitem[5]{5}
R.  Genzel, F.  Eisenhauer and S.  Gillessen:
\emph{The Galactic Center massive black hole and nuclear star cluster}.
Rev. Mod. Phys. \textbf{82} (2010),  3121--3196, arXiv: 1006.0064 [astro-ph. GA].

\bibitem[6]{6}
K. Gebhardt, J.  Adams, D.  Richstone, T.R.  Lauer et al:
\emph{The black-hole mass in M87 from GEMINI/NIFS adaptive optics observations}.
Astroph. J. \textbf{729} (2011), 119, arXiv: 1101.1954 [astro-ph.CO].


\bibitem[7]{7}
B.P.  Abbott et al. (LiGO Scientific Coll. and Virgo Coll.):~\emph{Observation of Gravitational Waves from a Binary Black Hole Merger}.
Phys. Rev. Lett. \textbf{116} (2016) 061102, arXiv:   1602.03837 [gr-qc]      

\bibitem[8]{8}
A.  Einstein:
\emph{N\"aherungsweise Integration der Feldgleichungen der Gravitation}.
Sitzungsber. Kgl. Preu{ss}. Akad. Wiss. zu Berlin, phys.-math. Klasse 1916, 688--696.

\bibitem[9]{9}
A.  Einstein:~\emph{\"Uber Gravitationswellen}.
Sitzungsber. Kgl. Preu{ss}. Akad. Wiss. zu Berlin, phys.-math. Klasse 1918, 154--167.


\bibitem[10]{10}
G.  Hinshaw, D.  Larson, E.  Komatsu, D.  Spergel et al.:
\emph{Nine-Year Wilkinson Microwave Anisotropy Probe (WMAP) Observations: Cosmological Parameter Results}.
Astrophys. J. Supplement \textbf{208} (2013), 19--44. arXiv: 1212.5226 [astro-ph.CO].

\bibitem[11]{11}
P.A.R.  Ade et al. (Planck Collaboration):
\emph{Planck 2015 results. XIII. Cosmological Parameters}, arXiv: 1502.01589 [astro-ph. Co].

\bibitem[12]{12}
\mbox{A.~Einstein:~\emph{Kosmologische~Betrachtungen~zur allgemeinen~Relativit\"atstheorie}}.
Sitzungsber. Kgl. Preu{ss}. Akad. Wiss. zu Berlin, phys.-math. Klasse 1917, 142--152.
    
\bibitem[13]{13}
J.  Levin, E.  Scannapieco and J.  Silk:
\emph{The topology of the universe: the biggest manifold of them all}.
Class. Quant. Grav. \textbf{15} (1998), 2689--2698, arXiv: gr-qc/9803026.

 \bibitem[14]{14}
 M.  Lachi\`eze-Rey and J.-P.  Luminet:
 \emph{Cosmic Topology}.
 Phys. Rep. \textbf{254} (1995), 135--214, arXiv: gr-qc/9605010.

\bibitem[15]{15}
N.J.  Cornish, D.N.  Spergel and G.D.  Starkman:
\emph{Circles in the sky: finding topology with the microwave background radiation}.
Class. Quant. Grav. \textbf{15} (1998),  2657--2670, arXiv:  astro-ph/9801212.

J.-P.  Luminet and  B.  Roukema: 
\emph{Topology of the Universe : Theory and Observation}.
In (M. Lachi\`eze-Rey , ed.) Theoretical and Observational Cosmology, Kluwer Acad. Publ., 1999, 117--157.

R.  Aurich and  F.  Steiner:
\emph{The cosmic microwave background for a nearly flat compact hyperbolic universe}.
Mon. Not. Roy. Astron. Soc. \textbf{323} (2001), 1016--1024, arXiv: astro-ph/0007264.

J.  Levin:
\emph{Topology and the cosmic microwave background}.
Phys. Rep. \textbf{365} (2002), 251-Ð333, arXiv: gr-qc/0108043.

R.  Aurich, S.  Lustig, F.  Steiner and  H.  Then:
\emph{Hyperbolic universes with a horned topology and the cosmic microwave background  anisotropy}.
Class. Quant. Grav. \textbf{21} (2004), 4901-4925, astro-ph/0403597.

J.-P.  Luminet, J.  Weeks, A.  Riazuelo, R.  Lehoucq and  J.-P.  Uzan:
\emph{Dodecahedral space topology as an explanation for weak wide-angle temperature correlations in the cosmic microwave background}.
Nature \textbf{425} (2003), 593--595, arXiv:astro-ph/0310253.

R.  Aurich, S.  Lustig and  F.  Steiner:
\emph{CMB Anisotropy of  the Poincar\'e Dodecahedron}.
Class. Quant. Grav. \textbf{22} (2005), 2061--2083,  arXiv: astro-ph/0412569.

R.  Aurich, S.  Lustig and  F.  Steiner:
\emph{CMB Anisotropy of Spherical Spaces}.
Class. Quant. Grav. \textbf{22} (2005), 3443--3460, arXiv: astro-ph/0504656.

M.J.  Rebou\c{c}as and  G.I.  Gomero:
\emph{Cosmic Topology: a Brief Overview}.
Braz. J. Phys. \textbf{34} (2004), 1358-1366, arXiv: astro-ph/0402324.

M.  Kunz,  N.  Aghanim, A.  Riazuelo and  O.  Forni:
\emph{On the detectability of non-trivial topologies}.
Phys. Rev. \textbf{D77} (2008) 023525, arXiv:  0704.3076 [astro-ph].

N. G.  Phillips and  A.  Kogut:
\emph{Constraints on the Topology of the Universe from the WMAP First-Year Sky Maps}.
Astroph. J. \textbf{645} (2006), 820--825, arXiv: astro-ph/0404400.

R.  Aurich,  H. S.  Janzer,  S.  Lustig and  F.  Steiner:
\emph{Do we live in a ''small universe''?}
 Class. Quant. Grav. \textbf{25} (2008) 125006, arXiv: 0708.1420.
 
P.M.  Vaudrevange, G.D.  Starkman, N.J.  Cornish and D.N.  Spergel :
\emph{Constraints on the Topology of the Universe: Extension to General Geometries}.
Phys. Rev. \textbf{D86} (2012) 083526, arXiv: 1206.2939 [astro-ph.CO].
 
B.F.  Roukema, M.J.  France, T.A.  Kazimierczak and  T.  Buchert:
\emph{Deep redshift topological lensing:  strategies for the $T^3$ candidate}.
Mon. Not. Roy. Astron. Soc. \textbf{437} (2014) 1096, arXiv: 1302.4425 [astro-ph.CO].

R.  Aurich:
\emph{A spatial-correlation analysis of the cubic 3-torus topology based on the Planck 2013 data}.
Mon. Not. Roy. Astron. Soc. \textbf{452} (2015), 1493-1501, arXiv: 1412.5355 [astro-ph.CO].

P.A.R.  Ade et al. (Planck Collaboration):
\emph{Planck 2015 results. XVIII. Background geometry and topology of the universe},
arXiv: 1502.01593 [astro-ph.CO].
 
\bibitem[16]{16}
P.  Appell:
\emph{Sur les fonctions de trois variables r\'eelles satisfaisant \`a l'\'equation $\Delta F=0$}.
Acta Mathematica, t.IV (1884), 313--374.


\bibitem[17]{17}
P.  Appel:
\emph{Sur quelques applications de la fonction $Z(x,y,z)$ \`a la Physique math\'ematique}.
Acta Mathematica, t.VIII (1886), 265--294.

\bibitem[18]{18}
P.  Appel:       
\emph{Sur les fonctions harmoniques \`a trois groupes de p\'eriodes}.
Rendiconti del circolo mathematico di Palermo, t.22 (1906), 361--370.  

\bibitem[19]{19}
P.  Epstein:~\emph{Zur Theorie allgemeiner Zetafunktionen}.
Math. Annalen \textbf{56} (1903), 615--644.

\bibitem[20]{20}
P.  Epstein:
\emph{Zur Theorie allgemeiner Zetafunktionen. II}.
Math. Annalen \textbf{63} (1907), 205--216.

 \bibitem[21]{21}
L.D.  Landau and E.M.  Lifschitz:
\emph{The Classical Theory of Fields}.
Pergamon Press, Oxford, 1962.

\bibitem[22]{22}
J.  Ehlers, I.  Ozsv\'ath  and  E.L.  Sch\"ucking:
\emph{Active Mass under Pressure}.
Am. J.  Phys.\textbf{74} (2006), 607--613, arXiv: gr-qc/0505040.

\bibitem[23]{23}
J.  Ehlers, I.  Ozsv\'ath ,  E.L.  Sch\"ucking and  Y.  Shang:
\emph{Pressure as a Source of Gravity}.
Phys. Rev. \textbf{D72} (2005) 124003, arXiv: gr-qc/0510041.

\bibitem[24]{24}
T.  Levi-Civit\'a:
\emph{Statica Einsteiniana}.
Rendiconti della R. Academia dei Lincei \textbf{26} (1917), 469.

\bibitem[25]{25}
R.C.  Tolman:
\emph{On the use of the energy-momentum principle in general relativity}.
Phys. Rev. \textbf{35} (1930), 875--895.

\bibitem[26]{26}
C.W.   Misner and  P.  Putnam:~\emph{Active gravitational mass}.
Phys. Rev. \textbf{116} (1959), 1045--1046.

\bibitem[27]{27}
B.  Schutz:
\emph{A First Course in General Relativity}.
Second Edition, Cambridge Univ. Press, 2009.

\bibitem[28]{28}
E.I.  Fredholm:
\emph{Sur une classe d'\'equations fonctionnelles}.
Acta Math. \textbf{27} (1903), 365--390.

\bibitem[29]{29}
R.  Courant and D.  Hilbert:
\emph{Methods of Mathematical Physics}.
Vol. I, Interscience Publ., New York (1953).

\bibitem[30]{30}
\mbox{T.~Tao:~\emph{A~proof~of~the~Fredholm~alternative}}.\\
https://terrytao.wordpress.com/2011/04/10.

\bibitem[31]{31}
Y.  Hagihara:
\emph{Theory of Relativistic Trajectories in a Gravitational Field of Schwarzschild}.
Japan J. Astronomy and Geophys. \textbf{8} (1931), 67--176.

\bibitem[32]{32}
E.  Hackmann and  C.  L\"ammerzahl:
\emph{\mbox{Geodesic equation in Schwarzschild-(anti-)} de Sitter space times: Analytical solutions and applications}.
Phys. Rev. \textbf{78} (2008) 024035, arXiv: 1505.07973 [gr-qc].

\bibitem[33]{33}
W.  Magnus, F.  Oberhettinger and  R.P.  Soni:
\emph{Formulas and Theorems for the Special Functions of Mathematical Physics}.
Third Edition, Springer-Verlag, Berlin, Heidelberg, New York (1966).

\bibitem[34]{34}
W.  Arendt, R.  Nittka, W.  Peter and  F.  Steiner:
\emph{Weyl's Law: Spectral Properties of the Laplacian in Mathematics and Physics}. 
In (eds. W.  Arendt, W. P.  Schleich),  Mathematical Analysis of Evolution, Information and Complexity, Wiley-VCH Verlag, Weinheim (2009), 1--71.

\bibitem[35]{35}
M.  Born:
\emph{On the stability of crystal lattices}. 
I. Math. Proc. Cambridge Philosophical Soc. \textbf{36} (1940), 160--172.

R.D.  Misra: 
\emph{On the stability of crystal lattices}.
II. ibid. \textbf{36} (1940), 173--182. 

M.  Born and R.D.  Misra:
\emph{On the stability of crystal lattices}.
IV. ibid. \textbf{36} (1940), 466--478.

\bibitem[36]{36}
B.F.  Roukema, S.  Bajtlik, M.  Biesiada,  A.  Szaniewska et al.:
\emph{A weak acceleration effect due to residual gravity in a multiply connected universe}.
Astronomy and Astroph. \textbf{463} (2007), 861--871, arXiv: astro-ph/0602159.

B.  Roukema and  P.T.  R\'o\.{z}a\'nski:
\emph{The residual gravity acceleration effect in the Poincar\'e dodecahedral space}.
Astronomy and Astroph.  \textbf{502} (2009),  27--38, arXiv: 0902.3402 [astro-ph.CO].

J.J. Ostrowski,  B.F.  Roukema and  Z.P.  Buli\'nski:
\emph{A relativistic model of the topological acceleration effect}.
Class. Quant. Grav. \textbf{29} (2012) 165006, 	arXiv: 1109.1596 [astro-ph.CO].
 
\bibitem[37]{37}
D.  Korotkin and  H.  Nicolai:
\emph{A Periodic Analog of the Schwarzschild Solution}.
DESY report TH94-038, arXiv: gr-qc/9403029.

\bibitem[38]{38}
G.D.  Birkhoff:
\emph{Relativity and Modern Physics}.
Harvard Univ. Press, Harvard, MA, 1923.

\bibitem[39]{39}
S.W. Hawking and  G.F.R.  Ellis:~\emph{The Large Scale Structure of Spacetime}.
Cambridge Univ. Press, Cambridge, 1973, Appendix B.

\bibitem[40]{40}
P.T.  Chru\'sciel, J.L.  Costa and  M.  Heusler:
\emph{Stationary Black Holes: Uniqueness and Beyond}.
Living Rev. Rel. \textbf{15} (2012), 1--69, arXiv: 1205.6112 [gr-qc].

\bibitem[41]{41}
S.W.  Hawking,  M.J.  Perry and  A.  Strominger:       
\emph{Soft Hair on Black Holes}.
Phys. Rev. Lett. \textbf{116} (2016), 231301, arXiv: 1601.00921 [hep-th].

\end{thebibliography}
\end{document}